\documentclass[journal]{IEEEtran}  

\ifCLASSINFOpdf

\else

\fi
\usepackage[linesnumbered,ruled,lined]{algorithm2e}
\usepackage{amssymb}

\usepackage{graphicx}
\usepackage{amsmath}
\usepackage{amssymb}
\usepackage{subfigure}
\usepackage{booktabs}
\usepackage{cite}
\usepackage{color}

\newtheorem{Rule}{Rule}

\newtheorem{theorem}{Theorem}
\newtheorem{definition}{Definition}

\newcommand{\eg}{\textit{e}.\textit{g}.}

\begin{document}

\title{Formation Control for Connected and Automated Vehicles on Multi-lane Roads:\\Relative Motion Planning and Conflict Resolution}

\author{Mengchi Cai$^{1}$,~\IEEEmembership{Student Member,~IEEE}, Qing Xu*$^{1}$, Chaoyi Chen$^{1}$,~\IEEEmembership{Student Member,~IEEE}, \\Jiawei Wang$^{1}$,~\IEEEmembership{Student Member,~IEEE}, Keqiang Li$^{1}$, Jianqiang Wang$^{1}$, Qianying Zhu$^{2}$
\thanks{This work was supported by the National Key Research and Development Program of China under Grant 2018YFE0204302, the National Natural Science Foundation of China under Grant 52072212, and Intel Collaborative Research Institute Intelligent and Automated Connected Vehicles.}
\thanks{$^{1}$Mengchi Cai, Qing Xu, Chaoyi Chen, Jiawei Wang, Keqiang Li and Jianqiang Wang are with School of Vehicle and Mobility, Tsinghua University, Beijing 100084, P.~R.~China.}
\thanks{$^{2}$Qianying Zhu is with Intel Lab China, Beijing 100084, P.~R.~China.}
\thanks{Corresponding author: Qing Xu, E-mail: qingxu@tsinghua.edu.cn}
}

\maketitle

\begin{abstract}

Multi-vehicle coordinated decision making and control can improve traffic efficiency while guaranteeing driving safety. Formation control is a typical multi-vehicle coordination method in the multi-lane scenario. Among the existing formation control methods, the formation switching process is predefined and the collision-free behavior of vehicles has not been considered and explained in detail. This paper proposes a formation control method for multiple Connected and Automated Vehicles (CAVs) on multi-lane roads. Firstly, a bi-level planning framework is proposed to switch the structure of the formation in different scenarios smoothly and effectively. Secondly, the relative coordinate system is established and the conflict-free relative paths are planned in the upper level. Then, multi-stage trajectory planning and tracking are modeled and solved as an optimal control problem with path constraints in the lower level. Next, case study is conducted to verify the function of the proposed method in different scenarios. Finally, simulation in the lane-drop bottleneck scenario is carried out under different traffic volume and numerical results indicate that the proposed method can improve both traffic efficiency and fuel economy at high traffic volume.

\end{abstract}


\IEEEpeerreviewmaketitle


%
\section{Introduction}
\label{intro}
%

Coordinated control for multiple Connected and Automated Vehicles (CAVs) has been proved to be able to improve driving safety of single vehicle and efficiency of the whole traffic system than single-vehicle automated control in multiple traffic scenarios~\cite{li2017dynamical, chen2020mixed, zheng2020smoothing}. Most of the existing multi-vehicle coordinated control methods focus on single-lane traffic scenarios where only the longitudinal behavior of vehicles is considered,~\eg\ platoon control on the single-lane road segments~\cite{bian2019reducing, naus2010string, zheng2016distributed}, and intersection coordination where each arm has only one approaching lane~\cite{xu2018distributed, li2006cooperative, xu2019cooperative}. Less research looks into the coordination of multiple vehicles on multi-lane traffic scenarios.

Lane assignment, on-ramp merging and lane-drop bottleneck merging are the three main research topics in the field of multi-lane coordination. Lane assignment methods aim to assign proper lanes to vehicles based on the lane capacity and occupancy of the highway system~\cite{dao2007optimized, hall1996optimized, dao2008distributed}. It helps distribute vehicles spatially on the drivable lanes to prevent or reduce traffic congestion. These methods provide optimal lane occupation suggestion for vehicles but not actually control them, which limits the improvement of traffic efficiency. On-ramp and lane-drop bottleneck merging control methods aim to resolve conflict for vehicles driving on different lanes. The difference of these two topics is that vehicles are allowed to change lanes at any collision-free points in lane-drop bottleneck scenarios but have to wait until the merging point at on-ramps. In the existing research, priority is usually designed and vehicles are controlled to drive according to the calculated passing order~\cite{hu2019trajectory, li2018consensus, xu2020bi, jing2019cooperative, phan2019space}. The aforementioned methods focus on solving the problem of traffic congestion within local scenarios. However, there still lacks a multi-vehicle coordination framework that provides driving strategies for vehicles to adapt to multiple traffic scenarios effectively and smoothly.

\begin{figure}
\begin{center}
    \subfigure[Formation switching at lane-drop bottleneck area]{
    \includegraphics[width=0.95\linewidth]{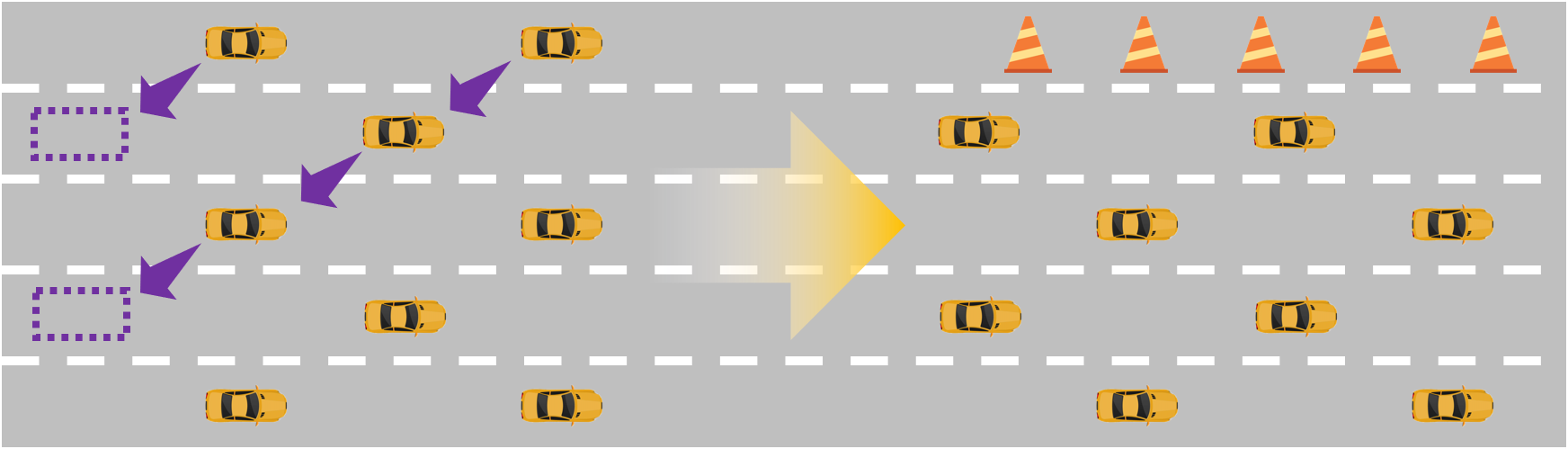}
    \label{formation1}}
    \subfigure[Formation switching for lane clearance]{
    \includegraphics[width=0.95\linewidth]{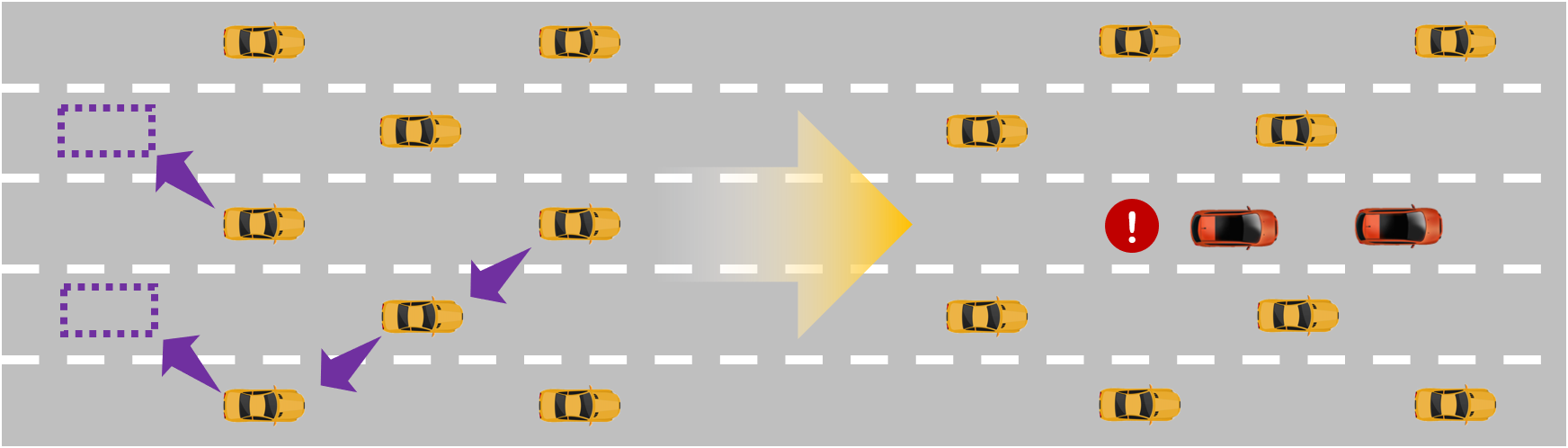}
    \label{formation2}}
    \caption{Typical scenarios for vehicular FC. The yellow vehicles are with FC and the red vehicles represent low-speed or emergency vehicles. The vehicular group with FC is able to switch structure adaptively according to the driving scenarios. }
    \label{formationexample}
\end{center}
\end{figure}

Formation control (FC), also called as convoy control, considers multiple CAVs as a group and plans the movement of vehicles from the overall perspective \cite{kato2002vehicle, marinescu2010active, marinescu2012ramp, marjovi2015distributed, cai2019multi, xu2021coordinated, jond2020autonomous, wu2020emergency, firoozi2021formation, navarro2016distributed, cao2021platoon}. The main difference between FC methods and the above on-ramp merging and lane-drop bottleneck coordination methods is that the FC methods focus on the global coordination of vehicles. Vehicles drive as a whole and switch the formation structure adaptively according to the changing driving environment, which enables the FC methods to be applied to multiple scenarios,~\eg\ lane-drop bottlenecks, on-ramps, off-ramps, and lane clearance for low-speed or emergency vehicles, as shown in Fig.~\ref{formationexample}. The idea of multi-lane FC of vehicles is firstly proposed in~\cite{kato2002vehicle}. The spatial distribution of vehicles in the formation is designed and the formation is able to cooperatively change shape to pass lane-drop bottlenecks smoothly. The slot-based approach is proposed and developed in~\cite{marinescu2010active, marinescu2012ramp} based on the coordination of vehicles and infrastructure. Moving slots are generated by the infrastructure and vehicles follow the slots to perform multi-lane driving, on-ramp merging and lane-drop bottleneck coordination. The distributed graph-based convoy control method for group of vehicles to maintain the formed formation is proposed in~\cite{marjovi2015distributed}, where distributed communication topology of vehicles is defined and the formation controller is designed. The idea of combining FC and task assignment for CAVs is firstly proposed in~\cite{cai2019multi}. Vehicles are assigned to generated targets in real time when the formation structure is changed according to the driving environment. This enables the formation to dynamically change its structure and improve its scenario adaptability, e.g., the FC method is proved to be able to improve traffic efficiency in multi-lane intersections~\cite{xu2021coordinated}. Other research focusing on vehicular FC includes~\cite{navarro2016distributed, jond2020autonomous, wu2020emergency, firoozi2021formation, cao2021platoon}. The shortcomings of the existing FC methods include that: (1) the formation switching process is predefined and the explanation of how the formation reacts to different scenarios is lacked; (2) the collision-free behavior of vehicles in formation switching process has not been considered and explained in detail.

This paper proposes a multi-lane FC method for multiple CAVs to improve the overall traffic efficiency and guarantee the smoothness of formation switching process between different scenarios. The main contributions of this paper are as follows:

\begin{enumerate}
\item A bi-level motion planning framework is proposed to guarantee safety and formation-switching smoothness of FC for CAVs. In the upper level, conflict-free relative path planning for vehicles is performed in relative coordinate system. In the lower level, vehicles are controlled to track the generated trajectories. This framework is able to switch the structure of formation smoothly and effectively according to the changing driving scenarios, because the motion of vehicles is regulated and planned cooperatively.
\item A conflict-free relative path planning algorithm is proposed in relative coordinate system, where motion conflicts of vehicles can be described more clearly than in the geodetic coordinate system, which is typical in existing research about vehicular formation control~\cite{kato2002vehicle, marinescu2010active, marinescu2012ramp, marjovi2015distributed, jond2020autonomous, wu2020emergency, firoozi2021formation, cao2021platoon}.  The conflict types of vehicles moving in formation are defined and potential collision is avoided by iteratively updating assignment result. 
\item Relative path map is built to describe the switching process, and detect and resolve the conflicts of vehicles. In the map, conflict-free relative motion sequences of vehicles are determined and passed to the lower level. In the lower level, the spatio-temporally constrained trajectory following problem is modeled and solved as an optimal control problem with path constraints.
\item Case study indicates that the proposed FC method using relative coordinate system is able to smoothly switch the structure of formation without collision in multiple traffic scenarios. Numerical results of the simulation in a lane-drop bottleneck scenario indicate that the proposed FC method can significantly improve both traffic efficiency and fuel economy of vehicles under high traffic volume.
\end{enumerate}

Some preliminary results have been presented in a conference version~\cite{cai2021formation}. The rest of this paper is organized as follows. Section~\ref{bilev} defines the bi-level motion planning framework and establishes the relative coordinate system. Section~\ref{moti} and Section~\ref{mult} provide the upper-level conflict-free motion planning and lower-level path-constrained trajectory planning methods respectively. Section~\ref{simu} carries out case study and simulations. Section~\ref{conc} gives the conclusion of this paper.

%
\section{Bi-level Motion Planning Framework and the Relative Coordinate System}
\label{bilev}
%

In this section, the bi-level motion planning framework is introduced. The relative coordinate system is built for relative path planning and motion regulation in the upper level, and multi-stage trajectory planning and tracking is performed in the lower level. 

\subsection{Bi-level Motion Planning Framework}

\begin{figure}
\begin{center}
    \includegraphics[scale = 0.2]{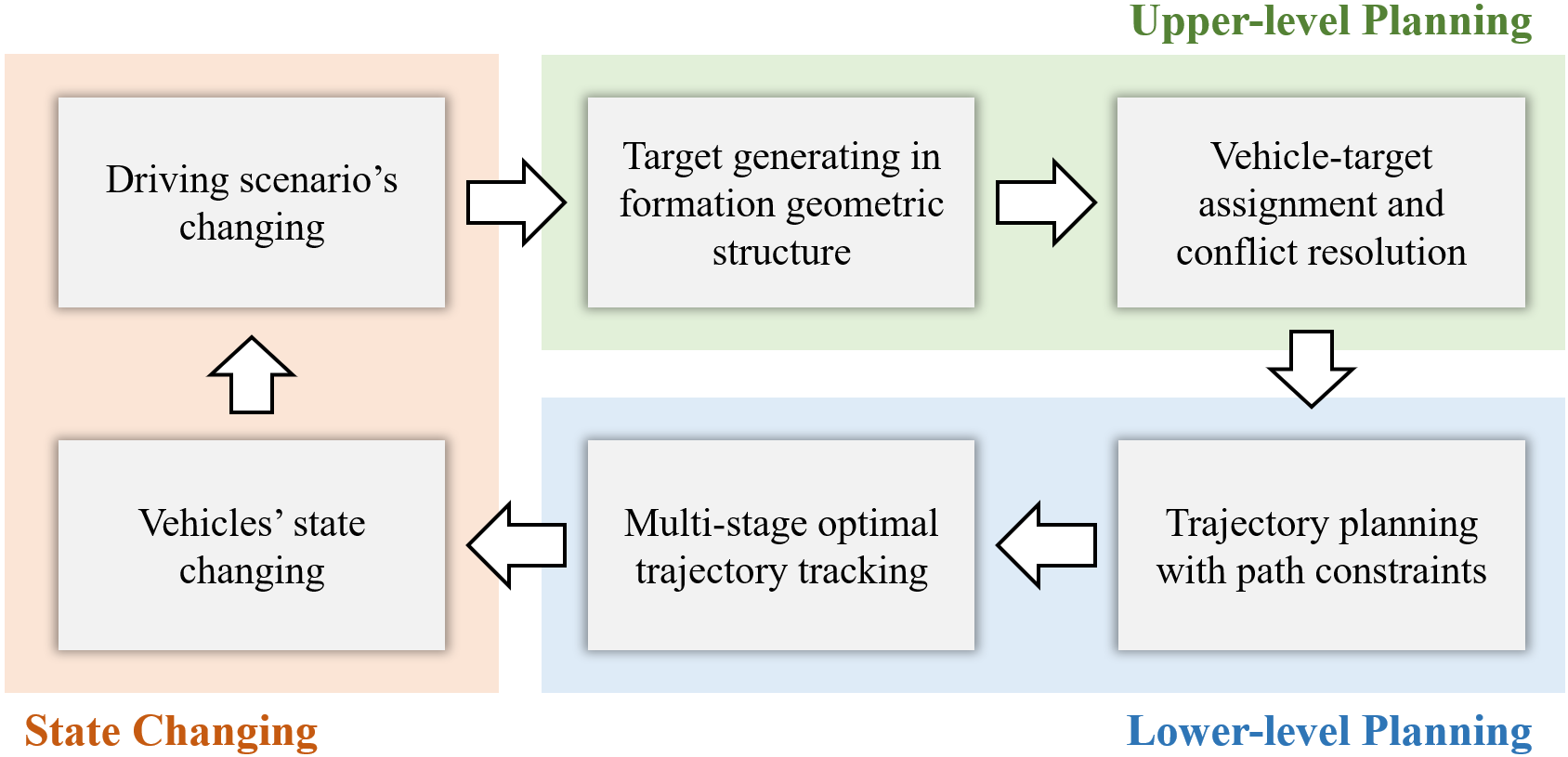}
    \caption{The architecture of decision making and control of this research. The orange part represents the state changing in the real world, consisting of states of vehicles and scenarios. The green part and the blue part represent the upper-level planning and the lower-level planning respectively.}
    \label{archi}
\end{center}
\end{figure}

In order to guarantee collision-free driving and formation switching smoothness, a bi-level motion planning framework is proposed in this paper. The architecture of the framework is shown in Fig.~\ref{archi}. 

When the driving scenario changes, the upper-level decision maker chooses the appropriate geometric structure and generates targets (Section~\ref{fgs}). Combining the information of vehicles' state and targets' position, the one-to-one matching between vehicles and targets is done  (Section~\ref{soap}) and conflict-free relative paths for vehicles to travel towards their targets are calculated  (Section~\ref{tdsaa} and Section~\ref{mscr}). The upper-level paths are then sent to the lower-level planners. Vehicles then perform trajectory planning with spatiotemporal constraints and optimal trajectory tracking  (Section~\ref{mult}). This process changes the state of vehicles to adapt to the changed driving environment.

\begin{figure}[t]
\begin{center}
    \includegraphics[scale = 0.23]{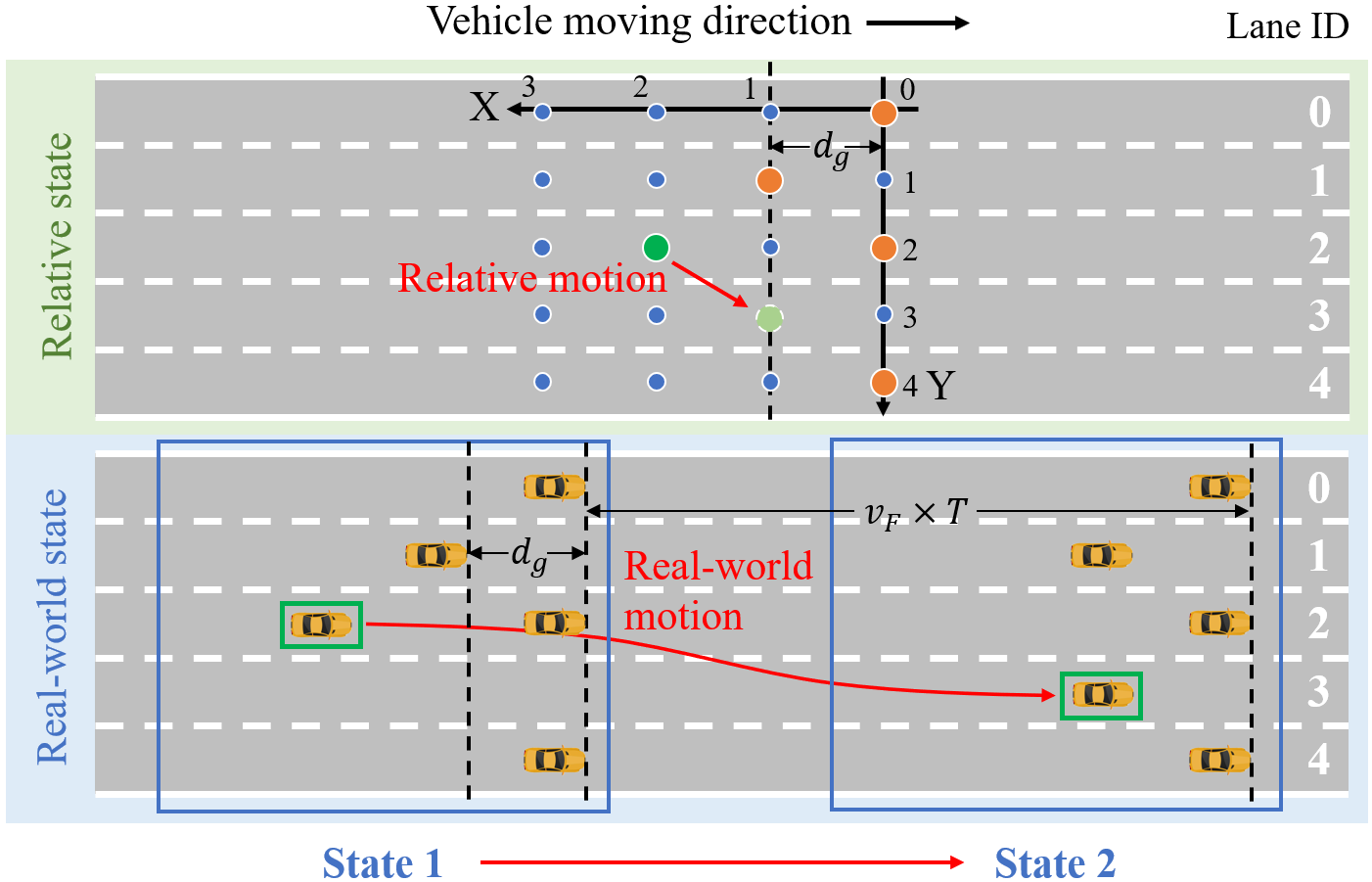}
    \caption{The bi-level motion planning framework of this research. The upper half contains the relative state and motion of vehicles, where blue and orange circles represent $\text{P}_{\text{re}}$s in RCS, and the green circles represent the target vehicle. The lower half represent the state and motion of vehicles in the real world. The vehicle in green rectangle is the target vehicle corresponding to the green circles in relative state. The points that vehicles occupy in `Real-world State' are $\text{P}_{\text{ro}}$s.}
    \label{bilevel}
\end{center}
\end{figure}

Fig.~\ref{bilevel} shows the working process of the bi-level motion planning framework. The upper-level planner firstly calculates collision-free paths for vehicles. The paths consist of several key points which can guide vehicles to avoid collision with others and switch the formation to the desired structure. In the lower level, vehicles generate and track trajectories to pass the key points calculated by the upper level. 

The conflicts and potential collisions of vehicles in the Geodetic Coordinate System (GCS) is hard to handle because of the nonlinear vehicle model. However, if we change the perspective from GCS to the moving vehicles, their relative state and motion can be described much more easily. Thus, we establish the Relative Coordinate System (RCS) to describe and regulate vehicles' relative motion and the vehicles can be treated as free-moving robots. The established RCS is shown in the `Relative state' in Fig.~\ref{bilevel}.

\subsection{Setup of RCS}

First of all, the perspective is transformed from GCS to the moving vehicles. According to the established relative axes, the most forward vehicle has the smallest $X$ coordinate. The $Y$ axis is set to pass through the center of rear axle of the most forward vehicle. The $X$ coordinate $x^\mathrm{r}$ of a vehicle represents the relative longitudinal distance $d^{\mathrm{r,lon}}$ between the vehicle and the most forward vehicle. The relative $X$ coordinate of vehicle $i$, noted as $x^\mathrm{r}_i$, can be expressed as:
\begin{eqnarray}
x^\mathrm{r}_i=d^{\mathrm{r,lon}}_{i,0}=d^{\mathrm{lon}}_{i,0}/d_\mathrm{g},
\end{eqnarray}
where $d^{\mathrm{r,lon}}_{i,0}$ represents the longitudinal relative distance between vehicle $i$ and the most forward vehicle in RCS, $d^{\mathrm{lon}}_{i,0}$ represents the longitudinal Euclidean distance between vehicle $i$ and the most forward vehicle in GCS, and $d_\mathrm{g}$ represents the safe following gap between the heads of two adjacent vehicles in the same lane.

The relative $Y$ coordinate $y^\mathrm{r}$ of a vehicle in RCS represents the ID of the lane that the vehicle occupies. The ordered pair $(x^\mathrm{r},y^\mathrm{r})$ represents the relative position of a vehicle in a formation, which is called as the Formation Relative Coordinate (FRC). The points in RCS whose $X$ coordinate $x^\mathrm{r}$ and $Y$ coordinate $y^\mathrm{r}$ are both integers are the key relative points ($\text{P}_{\text{re}}$s). Since RCS is a dynamic coordinate system that moves with vehicles in formations, the $\text{P}_{\text{re}}$s in RCS are one-to-one matched with key road points ($\text{P}_{\text{ro}}$s) in GCS at fixed time point.

\subsection{Motion Rules in RCS}
\label{theb}

In order to guarantee that no collision happens while vehicles move simultaneously in formation, some rules for relative motion in RCS are made.

\begin{Rule}
\label{rule1}
The time and space are both discretized for relative motion in RCS. The time is discretized with constant interval $T$, so that available time set in RCS is $\mathbb{T}=\{ t_i | t_i=iT, i=0,1,2,3...\}$, where $t_0$ is the initial time point. The space is discretized to $\text{P}_{\text{re}}$s. For each $t_i$ in $\mathbb{T}$ where $i>0$, the FRC of a vehicle must be at one of the $\text{P}_{\text{re}}$s.
\end{Rule}

\begin{Rule}
\label{rule2}
\label{RuleAdjacent}
For each two adjacent time points $t_i$ and $t_j$ in $\mathbb{T}$ where $i>0$ and $j=i+1$, the RCS of a vehicle $(x^\mathrm{r}_i,y^\mathrm{r}_i)$ and $(x^\mathrm{r}_j,y^\mathrm{r}_j)$ satisfy the following constraints:
\begin{eqnarray}
x^\mathrm{r}_j-x^\mathrm{r}_i,\ y^\mathrm{r}_j-y^\mathrm{r}_i \in \{-1,0,1\}.
\end{eqnarray}
\end{Rule}

Rule \ref{rule2} indicates that for a vehicle moving in RCS, the allowable relative position after one time interval is limited to the eight adjacent points and the current occupying point.

The relationship of $\text{P}_{\text{re}}$s and $\text{P}_{\text{ro}}$s is shown in Fig.~\ref{bilevel}. The green vehicle temps to join the four forward vehicles to form a standard five-lane interlaced formation. As shown in the relative state, the initial $\text{P}_{\text{re}}$ is $(2,2)$ and the final $\text{P}_{\text{re}}$ is $(1,3)$. The relative motion that the vehicle should take is shown as `Relative motion'. For each relative state in RCS, there is a corresponding real-world state in GCS at the same fixed time point. The formation is supposed to move with a constant velocity $v_\mathrm{F}$, so the longitudinal distance between two $\text{P}_{\text{ro}}$s corresponding to the same $\text{P}_{\text{re}}$ at two adjacent time point is $v_\mathrm{F}\times T$. The lower-level planner then performs trajectory planning for the vehicle to travel through the sequence of $\text{P}_{\text{ro}}$s.

%
\section{Upper-level Motion Planning}
\label{moti}
%

This section introduces the process of the upper-level motion planning. The relative motion planning for vehicles in RCS can be divided into three steps: choosing and generating of formation geometric structure, optimal assignment for vehicles and targets, and conflict resolution between vehicles.

\subsection{Formation Geometric Structure}
\label{fgs}

\begin{figure}
\begin{center}
    \subfigure[The interlaced structure]{
    \includegraphics[scale = 0.2]{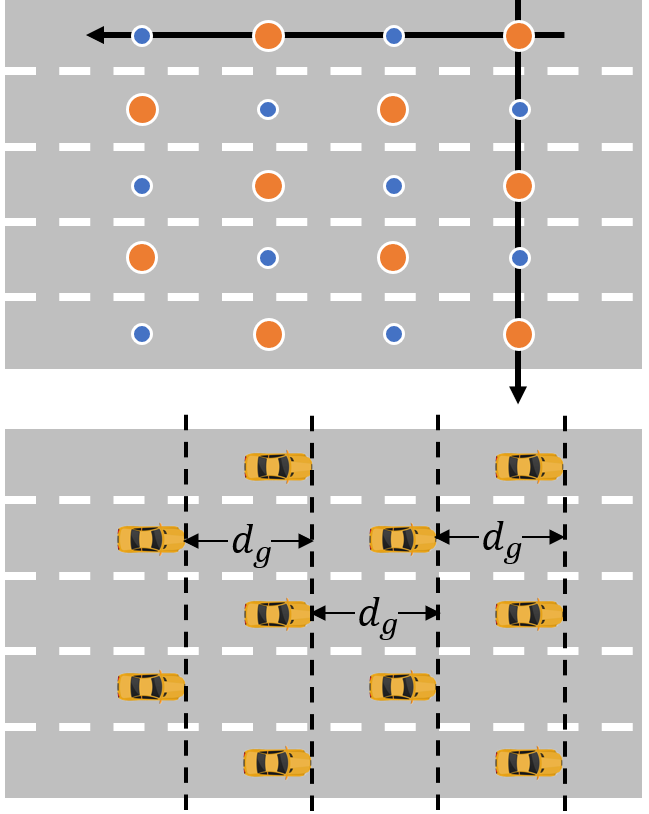}
    \label{interlaced}}
    \subfigure[The parallel structure]{
    \includegraphics[scale = 0.2]{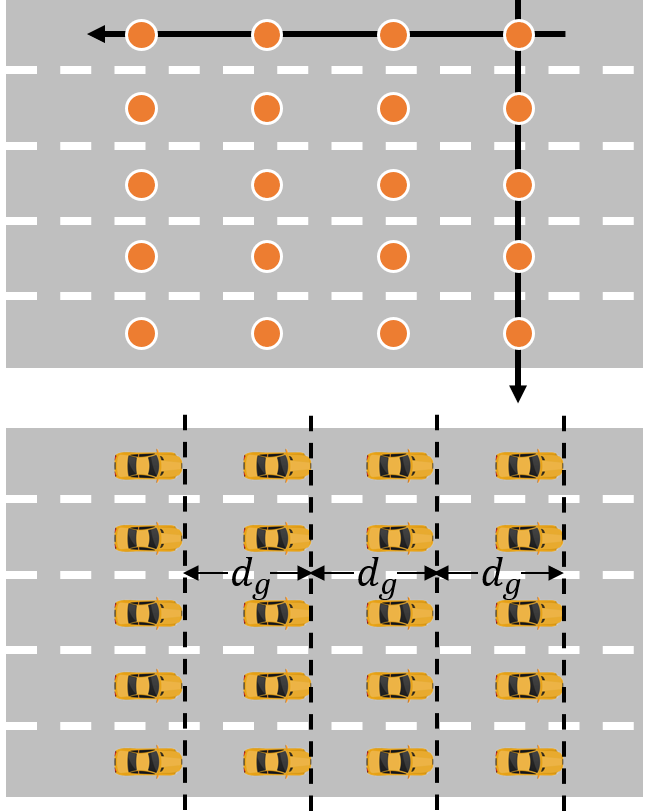}
    \label{parallel}}
    \caption{The two common formation geometric structure. The upper two figures show the occupancy of $\text{P}_{\text{re}}$s in RCS and the lower two figures show the real-world position of vehicles.}
    \label{formationstructure}
\end{center}
\end{figure}

The geometric structure of vehicular formations on multi-lane roads determines the relative position between vehicles. There are two common structure in the existing research: the interlaced structure (also known as diamond structure)~\cite{kato2002vehicle, marinescu2012ramp, marjovi2015distributed, cai2019multi, xu2021coordinated, jond2020autonomous} and the parallel structure~\cite{marinescu2010active, navarro2016distributed}, as shown in Fig. \ref{formationstructure}.

For the interlaced structure, vehicles leave some vacant space in the formation to make lane changing and formation switching more convenient. For the parallel structure, vehicles tend to occupy every available place and form tight formation. Existing research revealed that although the parallel structure has higher vehicle density, the interlaced structure is more suitable for multi-lane vehicle coordination considering lane-changing efficiency and driving safety~\cite{cai2019multi, xu2021coordinated, cao2021platoon}. In the rest of this paper, the interlaced structure is considered as the standard formation geometric structure. Additionally, vehicles should occupy more forward points in RCS prior to the other points.

\subsection{Vehicle-target Optimal Assignment}
\label{vtoa}

Targets are generated according to the geometric structure. The number of targets is equal to the number of vehicles that are going to form a formation. Then, an assignment needs to be done to one-to-one match vehicles and targets. 

\subsubsection{Setup of the Optimal Assignment Problem}
\label{soap}

An assignment builds the one-to-one matching relationship between vehicles and targets. A match between a vehicle and a target has a corresponding cost for the vehicle to travel to the target. The assignment with the minimum total matching cost of all the vehicles is the optimal assignment.

In order to find the optimal assignment, the cost to assign vehicle $i$ to target $j$ should be determined at first. Among the existing research, the Euclidean Distance (ED), the square of ED, the number of lane changes, or a weighted combination of the above indexes are often taken as the assigning cost~\cite{macdonald2011multi, cai2019multi, xu2021coordinated}. In this paper, since we solve the assignment problem in RCS, the Formation Relative Distance (FRD) is chosen as the assigning cost. The FRD $d^\mathrm{r}_{i,j}$ between the vehicle $i$ whose FRC is $(x^{\mathrm{r,v}}_i,y^{\mathrm{r,v}}_i)$ and the target $j$ whose FRC is $(x^{\mathrm{r,t}}_j,y^{\mathrm{r,t}}_j)$ is calculated as:
\begin{eqnarray}
\label{rd}
d^\mathrm{r}_{i,j}=&l^\mathrm{o}\times N^{\mathrm{r,o}}_{i,j}+l^\mathrm{s}\times N^{\mathrm{r,s}}_{i,j},\\
N^{\mathrm{r,o}}_{i,j}=&\text{min}(|x^{\mathrm{r,v}}_i-x^{\mathrm{r,t}}_j| , |y^{\mathrm{r,v}}_i-y^{\mathrm{r,t}}_j|),\\
N^{\mathrm{r,s}}_{i,j}=&\text{max}(|x^{\mathrm{r,v}}_i-x^{\mathrm{r,t}}_j| , |y^{\mathrm{r,v}}_i-y^{\mathrm{r,t}}_j|),\\
&-\text{min}(|x^{\mathrm{r,v}}_i-x^{\mathrm{r,t}}_j| , |y^{\mathrm{r,v}}_i-y^{\mathrm{r,t}}_j|),\notag
\end{eqnarray}
where $N^{\mathrm{r,o}}_{i,j}$ and $N^{\mathrm{r,s}}_{i,j}$ represent the number of the oblique and the straight edges of the path respectively, and $l^\mathrm{o}$ and $l^\mathrm{s}$ represent the weight of the oblique and the straight edges. An oblique edge connects two $\text{P}_{\text{re}}$s with different $X$ and $Y$ coordinates in RCS. In contrast, a straight edge connects two $\text{P}_{\text{re}}$s with the same $X$ or $Y$ coordinate. In the following of this paper, the following equation holds:
\begin{eqnarray}
l^\mathrm{s}=1,\ l^\mathrm{o}=1.
\end{eqnarray}

After determining the assignment cost between each vehicle and each target, the cost matrix $\mathcal{C}$, whose element on the $i$-th row and the $j$-th column represents the cost to assign vehicle $i$ to target $j$, can be calculated as:
\begin{eqnarray}
\mathcal{C}=[c_{i,j}]\in \mathbb{R}^{N\times N},\ c_{i,j}=d^\mathrm{r}_{i,j},\ i,j\in \mathbb{N}^+.
\end{eqnarray}

The assignment matrix $\mathcal{A}$, whose element on the $i$-th row and the $j$-th column represents whether vehicle $i$ is assigned to target $j$, is defined as:
\begin{eqnarray}
&\mathcal{A}=[a_{i,j}]\in \mathbb{R}^{N\times N}, \ 
\\&a_{i,j}=
\begin{cases}
1, \ \text{if vehicle $i$ is assigned to target $j$},\notag \\
0, \ \text{otherwise}.
\end{cases}
\end{eqnarray}

Then, the assignment problem can be modeled as a 0-1 integer programming problem:
\begin{alignat}{2}
\min\quad & \sum_{i=1}^N\sum_{j=1}^N (c_{i,j}\times a_{i,j}) &{}& \label{eqn - lp},\\
\mbox{s.t.}\quad
&\sum_{i=1}^N a_{i,j}=1,\notag \\
&\sum_{j=1}^N a_{i,j}=1,\notag \\
&i,j\in \mathbb{N}^+,\notag
\end{alignat}
where $N$ is the total number of vehicles, $[c_{i,j}]$ is the given cost matrix, and $[a_{i,j}]$ is the variable assignment matrix.

\subsubsection{The Conflict-free Assignment Algorithm}
\label{tdsaa}

The Hungarian Algorithm (HA), which is commonly used for multi-agent task assignment, is able to solve the assignment problem provided in (\ref{eqn - lp}), and returns the assignment with minimum total cost~\cite{19kuhn1955hungarian}. The A* algorithm is then used to calculate the optimal path with the smallest relative travelled distance for vehicles to drive to their targets~\cite{hart1968formal}.

The travelled distance of the path calculated by the A* algorithm is also equal to the FRD between the two points. The A* algorithm calculates the relative paths and the sets of $\text{P}_{\text{re}}$s that need to be passed orderly for vehicles in the formation. For any two vehicles whose relative paths overlap, a collision may potentially happen. Some types of conflict, like that the paths of two vehicles cross, can be avoided by HA~\cite{macdonald2011multi}. However, HA cannot prevent collision when the paths of vehicles overlap during the formation switching process. In order to resolve the collision, we need to define the types of conflict firstly.

\begin{figure}
\begin{center}
    \includegraphics[scale = 0.16]{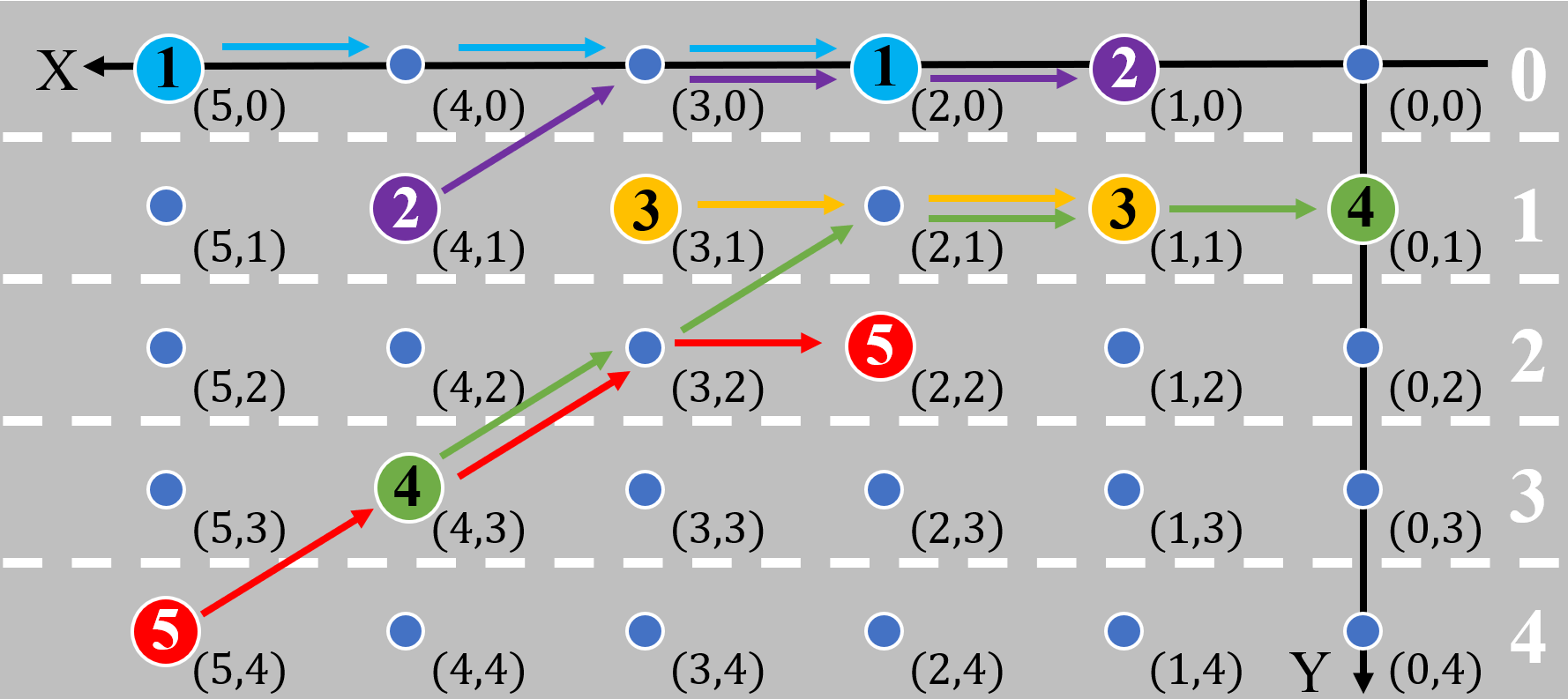}
\caption{The three defined conflict types. The bigger circles represent the starting and ending position of vehicles and the numbers in the circles are the ID of vehicles. The lines with arrows represent the relative paths of vehicles and each line is travelled in one time circle.}
\label{conflicttype}
\end{center}
\end{figure}

\begin{definition}
Define the types of conflict between two vehicles during the formation switching process: for vehicle $i$ and vehicle $j$, the targets assigned to them are target $k_i$ and target $k_j$ respectively. The conflict relationship can be categorized into three types:\\
Conflict Type 1: if target $k_i$ locates on the path for vehicle $j$ to travel to target $k_j$, and the steps for vehicle $i$ to travel to target $k_i$ is fewer than the steps for vehicle $j$ to travel to target $k_i$, which means:
\begin{eqnarray}
\begin{cases}
(x^{\mathrm{r,t}}_{k_i},y^{\mathrm{r,t}}_{k_i})\in\mathbb{P}^\mathrm{r}_j, \\
|\mathbb{P}^\mathrm{r}_i|<\max\{|x^{\mathrm{r,t}}_{k_i}-x^{\mathrm{r,v}}_j|,|y^{\mathrm{r,t}}_{k_i}-y^{\mathrm{r,v}}_j|\},
\end{cases}
\end{eqnarray}
then vehicle $i$ and vehicle $j$ are in Conflict Type 1, shown as vehicle 3 and 4 in Fig.~\ref{conflicttype};\\
Conflict Type 2: if target $k_i$ locates on the path for vehicle $j$ to travel to target $k_j$, and the steps for vehicle $i$ to travel to target $k_i$ is equal to or more than the steps for vehicle $j$ to travel to target $k_i$, which means:
\begin{eqnarray}
\begin{cases}
(x^{\mathrm{r,t}}_{k_i},y^{\mathrm{r,t}}_{k_i})\in\mathbb{P}^\mathrm{r}_j, \\
|\mathbb{P}^\mathrm{r}_i|\geq\max\{|x^{\mathrm{r,t}}_{k_i}-x^{\mathrm{r,v}}_j|,|y^{\mathrm{r,t}}_{k_i}-y^{\mathrm{r,v}}_j|\},
\end{cases}
\end{eqnarray}
then vehicle $i$ and vehicle $j$ are in Conflict Type 2, shown as vehicle 1 and 2 in Fig.~\ref{conflicttype};\\
Conflict Type 3: if vehicle $i$ and vehicle $j$ are not in Conflict Type 1 and Conflict Type 2, and part of their paths towards their assigned targets overlap, then they are in Conflict Type 3, shown as vehicle 4 and 5 in Fig.~\ref{conflicttype}.\\
\end{definition}

For any two vehicles where potential collision may happen, there must be some common $\text{P}_{\text{re}}$s of their relative paths $\mathbb{P}^\mathrm{r}$. However, not all the conflicts will cause collision. Among the three defined types of conflict, only the first type will cause deadlock and cannot be resolved by that one vehicle relatively stops and waits for the other to go. This will be explained in Section~\ref{mscr} in detail. However, the classical HA cannot resolve the conflict because the two possible assignments have the same total cost. In order to resolve the potential collision caused by the conflict, a novel algorithm, the conflict-free assignment algorithm, is proposed based on HA in Algorithm \ref{dsaa}. 
\begin{algorithm}[tb]
\label{dsaa}
\caption{The conflict-free assignment algorithm}  
\LinesNumbered  
\KwIn {$N^\mathrm{v}$: the number of vehicles\newline 
$\{(x^{\mathrm{r,v}}_i,y^{\mathrm{r,v}}_i)|i=1,2,3,...,N^\mathrm{v}\}$: FRC of vehicles.\newline 
$\{(x^{\mathrm{r,t}}_i,y^{\mathrm{r,t}}_i)|i=1,2,3,...,N^\mathrm{v}\}$: FRC of targets.\newline
$\mathcal{C}\in \mathbb{R}^{N^\mathrm{v}\times N^\mathrm{v}}$: the cost matrix.
}
\KwOut{$\mathcal{A}^\mathrm{H}\in \mathbb{R}^{N^\mathrm{v}\times 1}$: the optimal assignment result.\newline
$\{\mathbb{P}^\mathrm{r}_i|i=1,2,3,...,N^\mathrm{v}\}$: the relative paths for vehicles in RCS.
}  
\textbf{Step 1)} calculating the initial assignment $\mathcal{A}^\mathrm{H}$: apply HA to $N^\mathrm{v}$, $\{(x^{\mathrm{r,v}}_i,y^{\mathrm{r,v}}_i)\}$, $\{(x^{\mathrm{r,t}}_i,y^{\mathrm{r,t}}_i)\}$, and $\mathcal{C}$ to get $\mathcal{A}^\mathrm{H}$.\\
\textbf{Step 2)} relative path planning: \\
\For{$i\leftarrow 1$ \KwTo $N^\mathrm{v}$}{
set $k_i={\mathcal{A}^\mathrm{H}(i)}$.\\
apply A* algorithm to $(x^{\mathrm{r,v}}_i,y^{\mathrm{r,v}}_i)$ and $(x^{\mathrm{r,t}}_{k_i},y^{\mathrm{r,t}}_{k_i})$ to get $\mathbb{P}^\mathrm{r}_i$.
}
\textbf{Step 3)} conflict resolution:\\
\While{find any vehicles pairs $i$ and $j$ that are in Conflict Type 1}{
exchange the targets of vehicle $i$ and $j$ and update $\mathcal{A}^\mathrm{H}$.
}
\end{algorithm}

For the proposed  conflict-free assignment algorithm, the following theorems hold.

\begin{theorem}
\label{resproof}
Exchanging the targets of two vehicles can resolve the existing Conflict Type 1 and will not cause new Conflict Type 1.
\end{theorem}
\begin{IEEEproof}
The target position of vehicle $i$ divides the path of vehicle $j$ into two parts, $\mathbb{P}^\mathrm{r}_{j,1}$ and $\mathbb{P}^\mathrm{r}_{j,2}$. Suppose the travelled distance of $\mathbb{P}^\mathrm{r}_{i}$, $\mathbb{P}^\mathrm{r}_{j,1}$ and $\mathbb{P}^\mathrm{r}_{j,2}$ are $L_i$, $L_{j,1}$ and $L_{j,2}$, so the length of $\mathbb{P}^\mathrm{r}_{j}$ is $L_j=L_{j,1}+L_{j,2}$.
Vehicle $i$ will block the path of vehicle $j$ because $L_i<L_{j,1}$ and vehicle $i$ will arrive at the conflict point earlier than vehicle $j$. After exchanging the targets, the scenario changes to that the target of vehicle $j$ locates on the path of vehicle $i$ and vehicle $j$ will arrive at the point later than or at the same time with vehicle $i$, so the Conflict Type 1 is resolved and a new Conflict Type 2 is generated. Since Conflict Type 2 will not cause endless loop because the vehicle which is closer to its target can wait for a step to let the other vehicle pass (this will be discussed in section \ref{mscr}). The exchange of the two targets doesn't change the position of all the targets and the paths of other vehicles, so no new Conflict Type 1 will be generated. This completes the proof.
\end{IEEEproof}

\begin{theorem}
\label{finite}
Step 3 of Algorithm \ref{dsaa} will end in finite number of steps.
\end{theorem}
\begin{IEEEproof}
In Step 3 of Algorithm \ref{dsaa}, the loop will end if there is no Conflict Type 1 existing between any two vehicles. As is proved in Theorem \ref{resproof}, in each iteration of the loop, a  Conflict Type 1 will be resolved and no new  Conflict Type 1 will be generated. Since  Conflict Type 1 may exist between any two vehicles, when the number of vehicle is $N^\mathrm{v}$, the maximum number of  Conflict Type 1 is $(N^\mathrm{v}-1)+(N^\mathrm{v}-2)+(N^\mathrm{v}-3)+...+2+1=\frac{N^\mathrm{v}(N^\mathrm{v}-1)}{2}$, which is a finite number. After each iteration, the number of Conflict Type 1 will decrease by one, and the total number is finite, so the number will decrease into zero after finite iterations. This completes the proof.
\end{IEEEproof}

\begin{theorem}
\label{optproof}
Algorithm \ref{dsaa} returns the optimal assignment, which has the same total cost with the result of HA.
\end{theorem}
\begin{IEEEproof}
Since the classical HA is applied, the initial assignment $A^{\mathrm{H}}$ is optimal with the minimum cost. In each iteration of Algorithm \ref{dsaa}, the targets of two vehicles are exchanged and the A* algorithm is applied to calculate the new shortest paths.

We keep to consider the scenario in the proof of Theorem \ref{resproof}. When exchanging the targets of the two vehicles, the new calculated paths become $\mathbb{P}^{\mathrm{r}\prime}_{i}$ and $\mathbb{P}^{\mathrm{r}\prime}_{j}$, and the new length is calculated by:
\begin{eqnarray}
\begin{cases}
L'_i=L_i+L_{j,2},\\
L'_j=L_{j,1}.
\end{cases}
\end{eqnarray}
The total cost of the two vehicles is $L_t=L'_t=L_i+L_{j,1}+L_{j,2}$, so that the target exchange doesn't change the total cost of the assignment. This completes the proof.
\end{IEEEproof}

Theorem \ref{resproof} indicates that the assignment result will not cause unavoidable collision.  Theorem \ref{finite} guarantees that Algorithm \ref{dsaa} will return a solution and won't cause endless loop. Theorem \ref{optproof} guarantees the optimality of Algorithm \ref{dsaa}.

\subsection{Motion Sequences and Conflict Resolution}
\label{mscr}

Algorithm \ref{dsaa} calculates the optimal assignment for the initial position set of vehicles $\{(x^{\mathrm{r,v}}_i,y^{\mathrm{r,v}}_i)|i=1,2,3,...,N^\mathrm{v}\}$ and the generated set of targets $\{(x^{\mathrm{r,t}}_i,y^{\mathrm{r,t}}_i)|i=1,2,3,...,N^\mathrm{v}\}$. It also plans the relative paths $\{\mathbb{P}^\mathrm{r}_i|i=1,2,3,...,N^\mathrm{v}\}$ for the vehicles to travel to their assigned targets. Set $N^\mathrm{s}$ to the size of the longest path in $\{\mathbb{P}^\mathrm{r}_i\}$, and put the ID, the initial $\text{P}_{\text{re}}$, and the relative path of vehicle $i$ on the $i$-th row, we can get the relative path map for the $N^\mathrm{v}$ vehicles, as is shown in Table \ref{rpm}. For the vehicle whose path is shorter than $N^\mathrm{s}$, we can iterate the target position to the end of the path until it reaches the length $N^\mathrm{s}$.

\begin{table*}[htbp]
\begin{center}
\caption{Relative path map}
\label{rpm}
\begin{tabular}{cccccccccccc}
\toprule
Vehicle ID                &   Initial $\text{P}_{\text{re}}$       &step 1  &step 2 & ...   &step $j$& ...&step $N^\mathrm{s}$\\
\midrule
1&$(x^{\mathrm{r,v}}_{1,0},y^{\mathrm{r,v}}_{1,0})$     &   $(x^{\mathrm{r,v}}_{1,1},y^{\mathrm{r,v}}_{1,1})$    & $(x^{\mathrm{r,v}}_{1,2},y^{\mathrm{r,v}}_{1,2})$& ... & $(x^{\mathrm{r,v}}_{1,j},y^{\mathrm{r,v}}_{1,j})$& ...& $(x^{\mathrm{r,v}}_{1,N^\mathrm{s}},y^{\mathrm{r,v}}_{1,N^\mathrm{s}})$\\ 
2&$(x^{\mathrm{r,v}}_{2,0},y^{\mathrm{r,v}}_{2,0})$     &   $(x^{\mathrm{r,v}}_{2,1},y^{\mathrm{r,v}}_{2,1})$   & $(x^{\mathrm{r,v}}_{2,2},y^{\mathrm{r,v}}_{2,2})$& ... & $(x^{\mathrm{r,v}}_{2,j},y^{\mathrm{r,v}}_{2,j})$& ...& $(x^{\mathrm{r,v}}_{2,N^\mathrm{s}},y^{\mathrm{r,v}}_{2,N^\mathrm{s}})$\\ 
...            & ...      & ...                          & ...        &   ...      & ...  & ...      & ... \\ 
$i$&$(x^{\mathrm{r,v}}_{i,0},y^{\mathrm{r,v}}_{i,0})$        &   $(x^{\mathrm{r,v}}_{i,1},y^{\mathrm{r,v}}_{i,1})$     & $(x^{\mathrm{r,v}}_{i,2},y^{\mathrm{r,v}}_{i,2})$& ... & $(x^{\mathrm{r,v}}_{i,j},y^{\mathrm{r,v}}_{i,j})$& ...& $(x^{\mathrm{r,v}}_{i,N^\mathrm{s}},y^{\mathrm{r,v}}_{i,N^\mathrm{s}})$\\ 
...           & ...        & ...                    & ...      &   ...       & ...    & ...   & ... \\ 
$N^\mathrm{v}$&$(x^{\mathrm{r,v}}_{N^\mathrm{v},0},y^{\mathrm{r,v}}_{N^\mathrm{v},0})$ &$(x^{\mathrm{r,v}}_{N^\mathrm{v},1},y^{\mathrm{r,v}}_{N^\mathrm{v},1})$&$(x^{\mathrm{r,v}}_{N^\mathrm{v},2},y^{\mathrm{r,v}}_{N^\mathrm{v},2})$& ...& $(x^{\mathrm{r,v}}_{N^\mathrm{v},j},y^{\mathrm{r,v}}_{N^\mathrm{v},j})$& ...& $(x^{\mathrm{r,v}}_{N^\mathrm{v},N^\mathrm{s}},y^{\mathrm{r,v}}_{N^\mathrm{v},N^\mathrm{s}})$\\
\bottomrule  
\end{tabular}
\end{center}
\end{table*}

Table \ref{rpm} shows the steps that each vehicle will take to travel to its assigned target, which makes it clear to find and resolve the collision between vehicles during the formation switching process. If two vehicles arrive at the same $\text{P}_{\text{re}}$ at the same step, they will collide. For this case, the vehicle with the lower priority will add a step to maintain the last relative position before the collision point to avoid colliding with the vehicle with the higher priority. If this addition produces redundant step at the end, the last step will be removed. If this addition produces necessary step and enlarges $N^\mathrm{s}$, other vehicles should also add steps to maintain the same total length of the paths. The priority is set higher for the vehicles that are farther to their targets in order to avoid deadlock. This process will resolve Conflict Type 2.

In order to better describe how the relative path map and Algorithm \ref{dsaa} helps resolve the conflict between vehicles during the formation switching process, two example cases are designed. 

\begin{figure*}
\begin{center}
    \subfigure[Initial relative position]{
    \includegraphics[scale = 0.12]{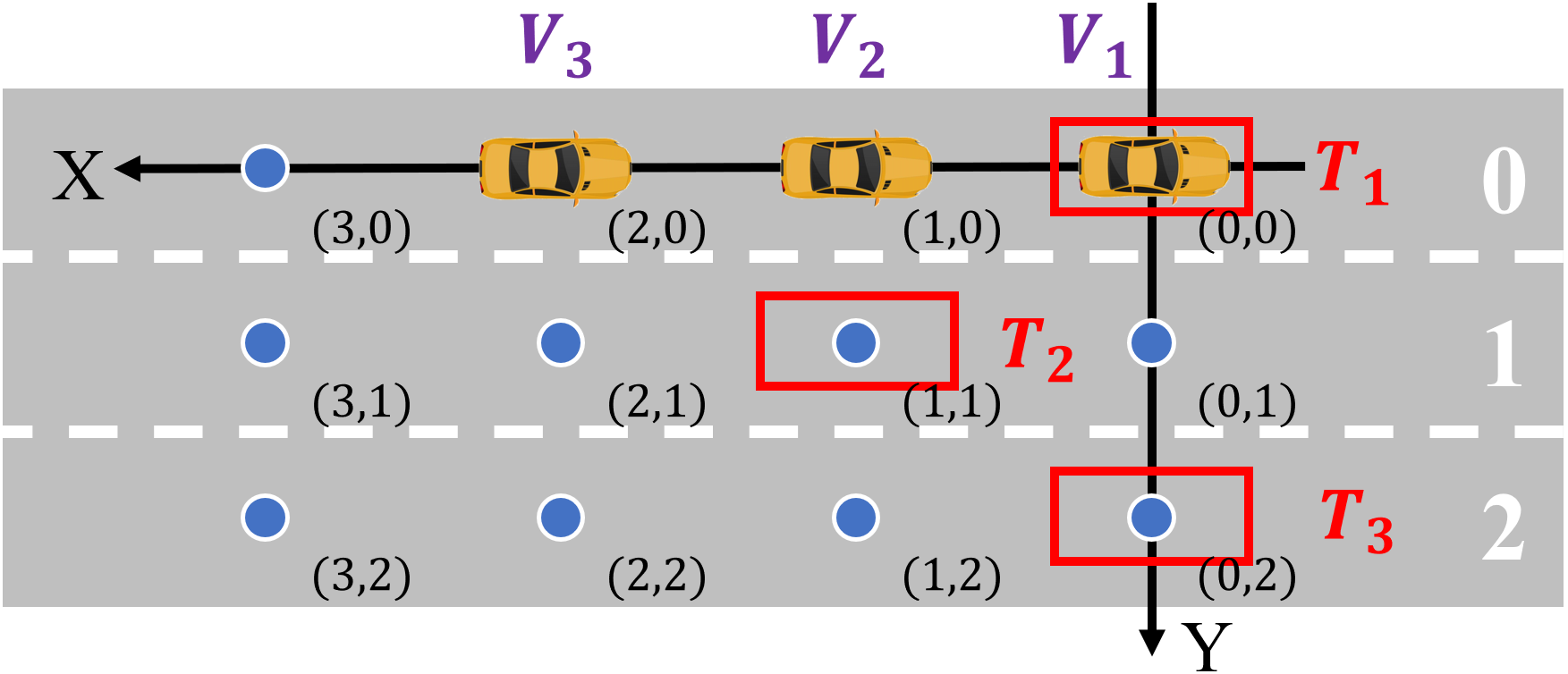}
    \label{case1_1}}
\hspace{0mm}
    \subfigure[Relative position at the $1^{st}$ step]{
    \includegraphics[scale = 0.12]{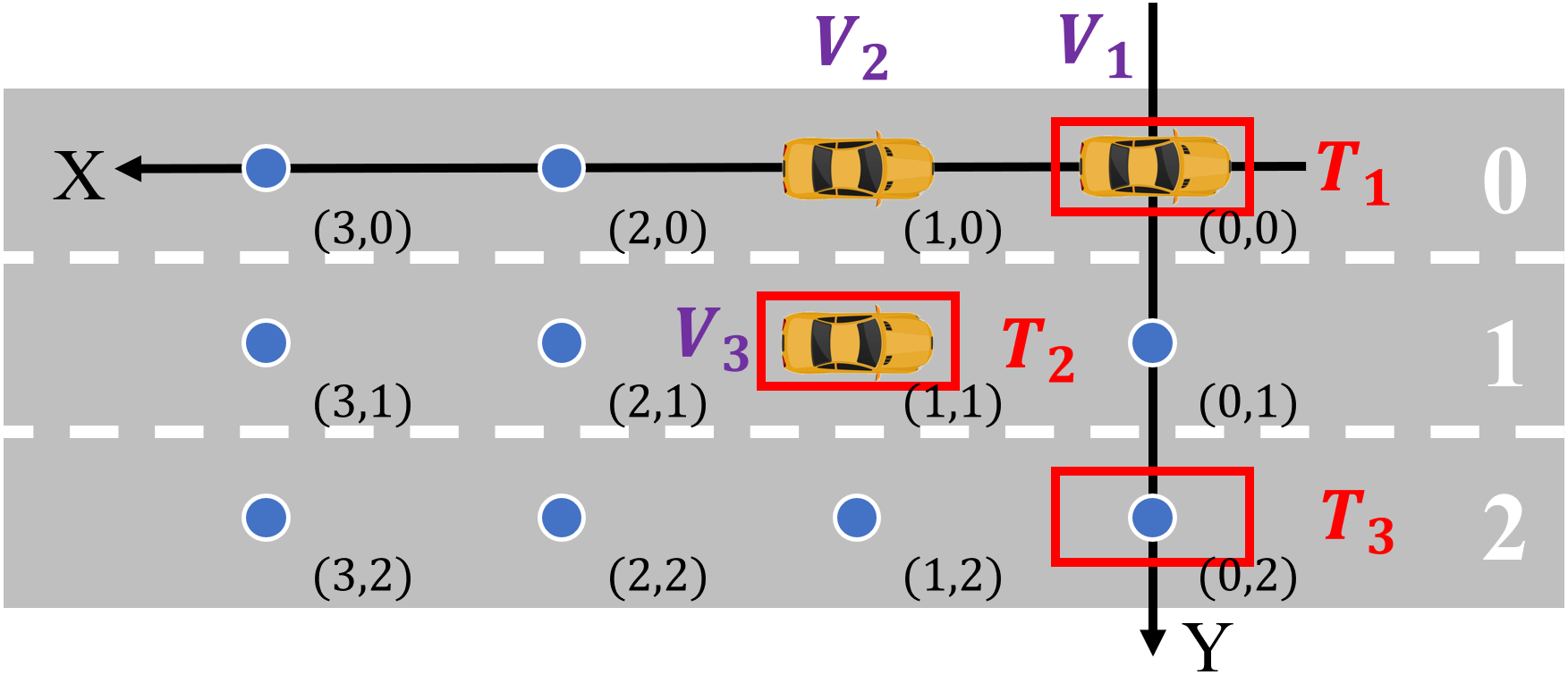}
    \label{case1_2}}
\hspace{0mm}
    \subfigure[Final relative position]{
    \includegraphics[scale = 0.12]{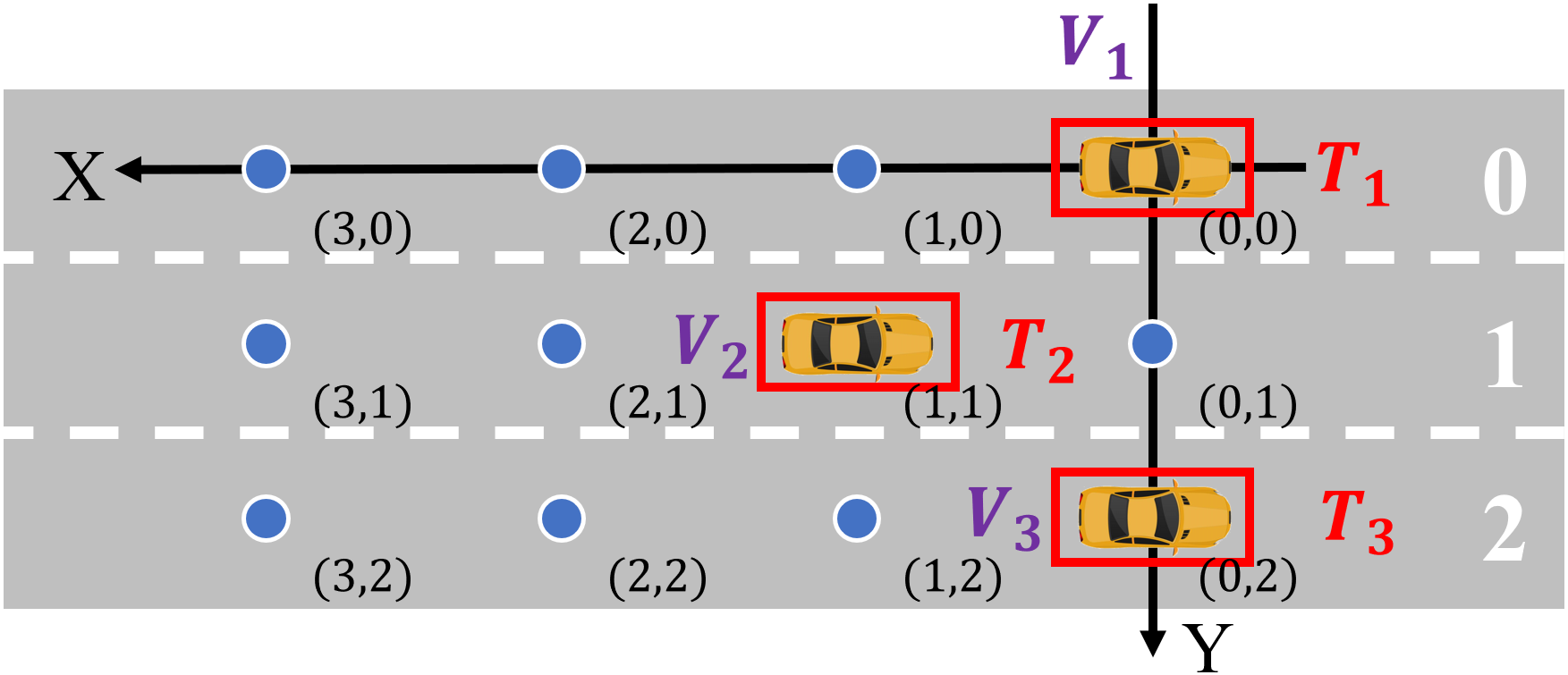}
    \label{case1_3}}
\hspace{0mm}
\caption{The switching process of case 1. The red rectangles represent targets. Vehicles and targets with same index of ID are one-to-one matched. In this case, vehicle 2 stays at its initial relative position for one time cycle to let vehicle 3 pass firstly, in order to resolve the conflict.}
\label{case1}
\end{center}
\end{figure*}

In case 1, which is shown in Fig. \ref{case1}, there are three vehicles, marked as $V_1$, $V_2$, and $V_3$, to be assigned to three targets marked as $T_1$, $T_2,$ and $T_3$. The assignment matrix calculated by HA is given in Table \ref{arfec}. Because no Conflict Type~1 exists in the result, the matrixes calculated by HA and Algorithm~\ref{dsaa} are the same. The initial geometric structure of the three-vehicle group is a one-lane platoon, and they need to switch to a three-lane formation. The A* algorithm is applied to calculate the relative paths for the vehicles, and the result is shown as Map~1 in Table \ref{rpm1}. Necessary steps are added to maintain the same total length, shown as Map~2.  From the map we can see that $V_2$ and $V_3$ are in Conflict Type~2. Because $V_3$ is farther away from its target than $V_2$, the priority of $V_3$ is set higher than $V_2$, and $V_2$ takes an additional step to maintain its initial relative position at step 1 and the length of the path of $V_2$ is increased by one. Since $V_2$ will reach its target at step~2, so that the additional step 3 of $V_2$ is redundant and thus removed, shown as Map~3. Map~4 shows the final calculated relative path map for the three vehicles, where no collision would happen and the formation switching process will end in two steps.

\begin{table}[htbp]
\begin{center}
\caption{Assignment results for example cases}
\label{arfec}
\begin{tabular}{cccccccccccc}
\toprule
Assignment matrix     &   Assignment matrix &    Assignment matrix \\
for case 1     &   for case 2 &    for case 2 \\
(HA)  &  (HA)  &  (Algorithm \ref{dsaa})\\
\midrule
$\begin{bmatrix}1& & \\ &1& \\ & &1\end{bmatrix}$  
&
$\begin{bmatrix} 1& & & \\ &1& & \\ & & 1&\\& & &1 \end{bmatrix}$ 
&
$\begin{bmatrix} & & &1 \\ &1& & \\ & & 1&\\1& & &\end{bmatrix}$   \\
\bottomrule
\end{tabular}
\end{center}
\end{table}

\begin{table}[htbp]
\begin{center}
\caption{The relative path maps for case 1}
\label{rpm1}
\begin{tabular}{cccccccccccc}
\toprule
Map 1 (HA)           \\
$
\begin{tabular}{cccccccccccc}
\toprule
Vehicle ID   &   Initial $\text{P}_{\text{re}}$       &step 1  &step 2   \\
\midrule
1&(0,0)\\
2&(1,0)&(1,1)\\
3&(2,0)&(1,1)&(0,2)\\
\bottomrule  
\end{tabular}
$\\
\midrule
 Map 2 (HA)\\
$
\begin{tabular}{cccccccccccc}
\toprule
Vehicle ID   &   Initial $\text{P}_{\text{re}}$       &step 1  &step 2   \\
\midrule
1&(0,0)&(0,0)&(0,0)\\
2&(1,0)&(1,1)&(1,1)\\
3&(2,0)&(1,1)&(0,2)\\
\bottomrule  
\end{tabular}
$\\
\midrule
Map 3 (HA)    \\
$
\begin{tabular}{cccccccccccc}
\toprule
Vehicle ID   &   Initial $\text{P}_{\text{re}}$       &step 1  &step 2   &step 3 \\
\midrule
1&(0,0)&(0,0)&(0,0)\\
2&(1,0)&(1,0)&(1,1)&(1,1)\\
3&(2,0)&(1,1)&(0,2)\\
\bottomrule  
\end{tabular}
$\\
\midrule
Map 4 (HA)\\
$
\begin{tabular}{cccccccccccc}
\toprule
Vehicle ID   &   Initial $\text{P}_{\text{re}}$       &step 1  &step 2    \\
\midrule
1&(0,0)&(0,0)&(0,0)\\
2&(1,0)&(1,0)&(1,1)\\
3&(2,0)&(1,1)&(0,2)\\
\bottomrule  
\end{tabular}
$\\
\bottomrule  
\end{tabular}
\end{center}
\end{table}

\begin{figure*}
\begin{center}
    \subfigure[Initial relative position]{
    \includegraphics[scale = 0.12]{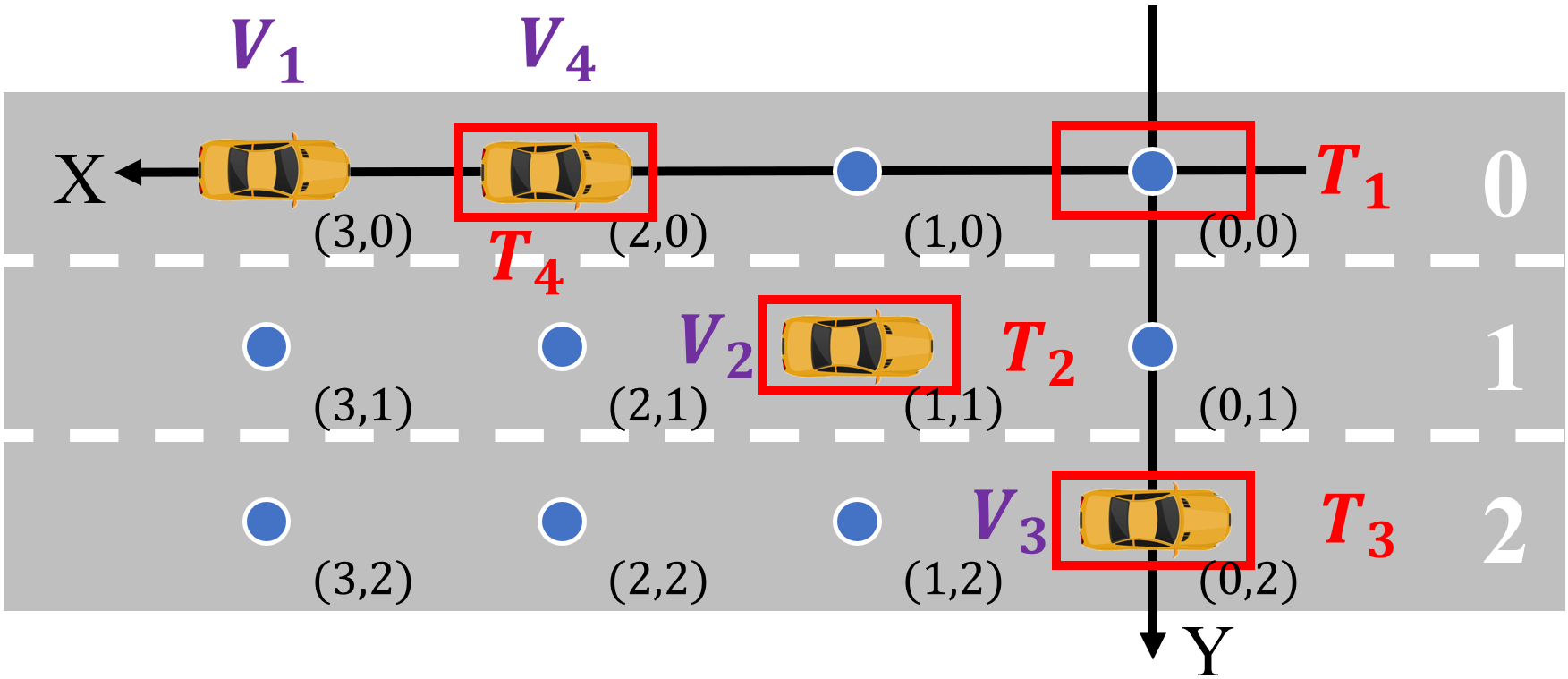}
    \label{case2_1}}
\hspace{0mm}
    \subfigure[Relative position at the $1^{st}$ step]{
    \includegraphics[scale = 0.12]{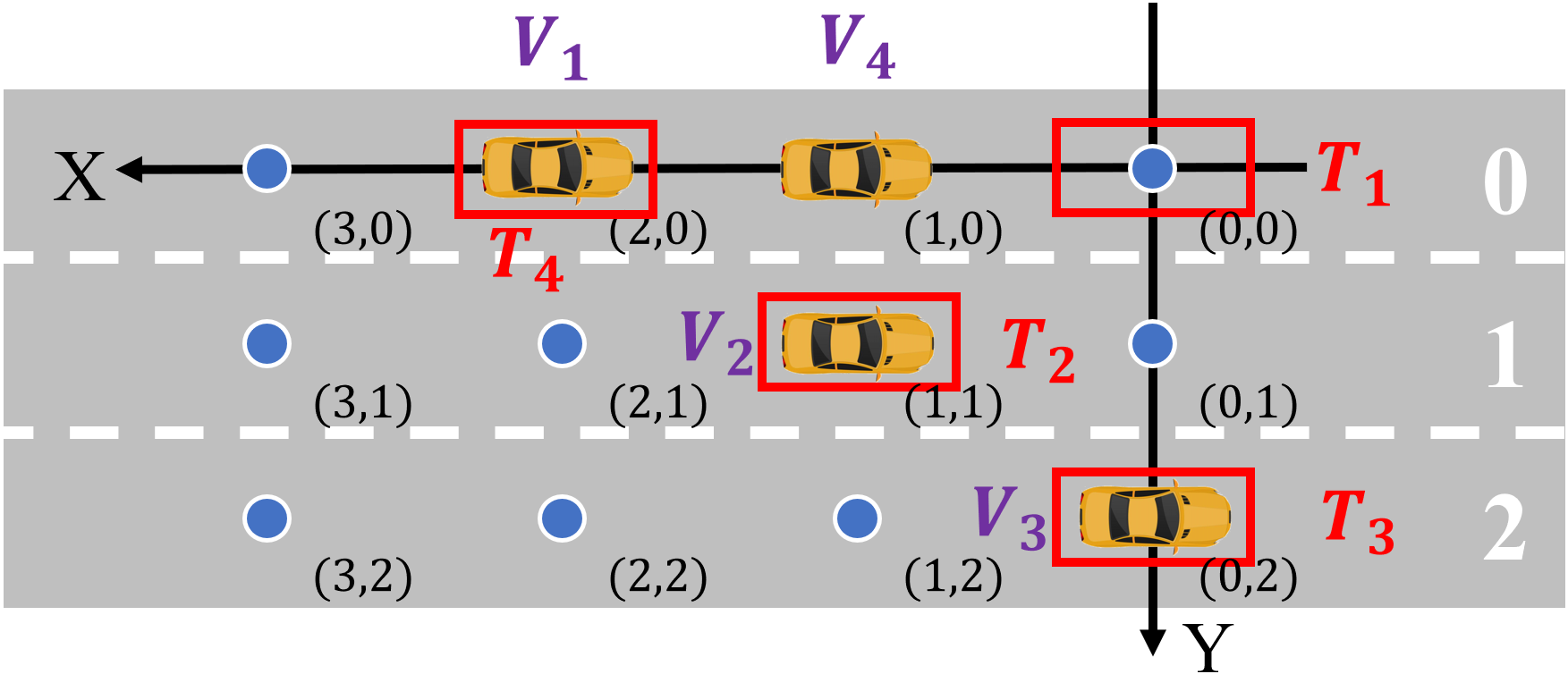}
    \label{case2_2}}
\hspace{0mm}
    \subfigure[Final relative position]{
    \includegraphics[scale = 0.12]{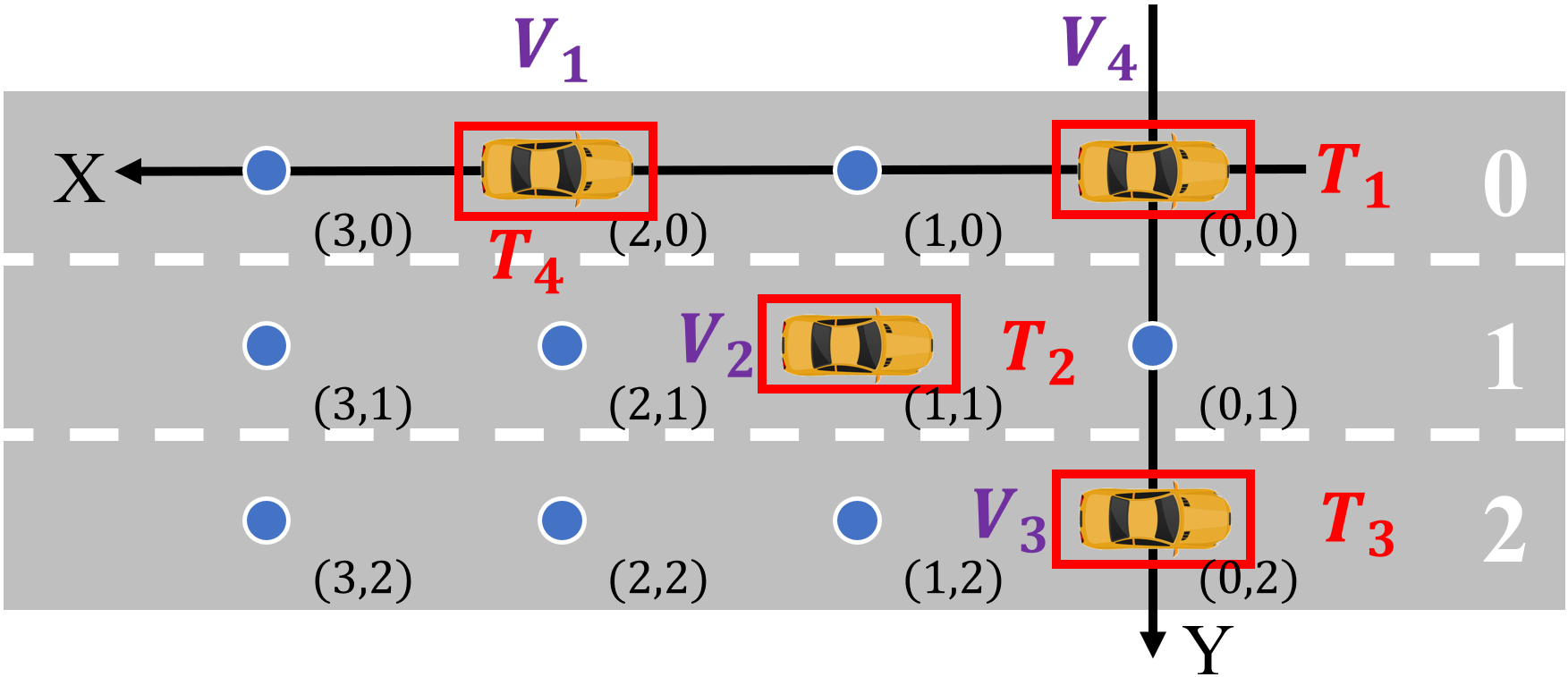}
    \label{case2_3}}
\hspace{0mm}
\caption{The switching process of case 2. Vehicles and targets with same index of ID are one-to-one matched initially and vehicle 4 and vehicle 1 are in Conflict Type 1. In order to resolve the conflict, Algorithm~\ref{dsaa} is adopted and the targets of vehicle 1 and vehicle 4 are exchanged.}
\label{case2}
\end{center}
\end{figure*}

In case 2, which is shown in Fig. \ref{case2}, there are four vehicles, marked as $V_1$, $V_2$, $V_3$ and $V_4$, to be assigned to four targets marked as $T_1$, $T_2$, $T_3$ and $T_4$. The assignment matrix calculated by HA is given in the $2^{nd}$ column in Table~\ref{arfec}, and the corresponding initial and stuffed relative path maps are shown as Map~1 (HA) and Map~2 (HA) in Table~\ref{rpm2}. Fig.~\ref{case2} shows that the relationship of $V_4$ and $V_1$ satisfies conflict type~1. Map~2 indicates that maintaining the last $\text{P}_{\text{re}}$ can't resolve the conflict and will cause endless loop, because $V_4$ will always block the path from $V_1$ to $T_1$. Thus, HA is not able calculate feasible solution for this case, and Algorithm~\ref{dsaa} is applied, whose result is shown in the $3^{rd}$ column in Table~\ref{arfec}. The corresponding initial and stuffed relative path maps are shown as Map~1 (Algorithm~\ref{dsaa}) and Map~2 (Algorithm~\ref{dsaa}) in Table~\ref{rpm2}, and it is clearly shown is Map~2 (Algorithm~\ref{dsaa}) that no collision would happen and the formation switching process will end in two steps.

\begin{table}[htbp]
\begin{center}
\caption{The relative path maps for case 2}
\label{rpm2}
\begin{tabular}{cccccccccccc}
\toprule
Map 1 (HA)            \\
$
\begin{tabular}{cccccccccccc}
\toprule
Vehicle ID   &   Initial $\text{P}_{\text{re}}$       &step 1  &step 2  &step 3 \\
\midrule
1&(3,0)&(2,0)&(1,0)&(0,0)\\
2&(1,1)\\
3&(0,2)\\
4&(2,0)\\
\bottomrule  
\end{tabular}
$\\
\midrule
 Map 2 (HA) \\
$
\begin{tabular}{cccccccccccc}
\toprule
Vehicle ID   &   Initial $\text{P}_{\text{re}}$       &step 1  &step 2  &step 3 \\
\midrule
1&(3,0)&(2,0)&(1,0)&(0,0)\\
2&(1,1)&(1,1)&(1,1)&(1,1)\\
3&(0,2)&(0,2)&(0,2)&(0,2)\\
4&(2,0)&(2,0)&(2,0)&(2,0)\\
\bottomrule  
\end{tabular}
$\\
\midrule
Map 1 (Algorithm \ref{dsaa})           \\
$
\begin{tabular}{cccccccccccc}
\toprule
Vehicle ID   &   Initial $\text{P}_{\text{re}}$       &step 1  &step 2   \\
\midrule
1&(3,0)&(2,0)\\
2&(1,1)\\
3&(0,2)\\
4&(2,0)&(1,0)&(0,0)\\
\bottomrule  
\end{tabular}
$\\
\midrule
Map 2 (Algorithm \ref{dsaa})\\
$
\begin{tabular}{cccccccccccc}
\toprule
Vehicle ID   &   Initial $\text{P}_{\text{re}}$       &step 1  &step 2  \\
\midrule
1&(3,0)&(2,0)&(2,0)\\
2&(1,1)&(1,1)&(1,1)\\
3&(0,2)&(0,2)&(0,2)\\
4&(2,0)&(1,0)&(0,0)\\
\bottomrule  
\end{tabular}
$\\
\bottomrule  
\end{tabular}
\end{center}
\end{table}

%
\section{Lower-level Trajectory Planning and Tracking}
\label{mult}
%

Motion planning in RCS calculates the collision-free relative paths for vehicles to travel to the assigned targets. The output of the relative motion planning is sequences of $\text{P}_{\text{re}}$s. In this section, the real-world trajectories which pass the $\text{P}_{\text{ro}}$s corresponding to the $\text{P}_{\text{re}}$s are generated for vehicles. Then, the multi-stage motion control method is proposed to reduce the tracking error.

\subsection{Trajectory Planning Using B$\acute{\text{e}}$zier Curves}

B$\acute{\text{e}}$zier curves are commonly used for trajectory planning for CAVs~\cite{choi2008path, han2010bezier}. Generated according to control points, the curve starts from the first control point and end at the last control point. Other control points determine the shape and curvature of the curve. The B$\acute{\text{e}}$zier curve is tangent to the line of the first two control points at the starting point and the line of the last two control points at the ending point, which make it a good choice for curvature-continuous trajectory planning for autonomous vehicles. Since there are possibly more than two $\text{P}_{\text{ro}}$s for the vehicles to pass, the whole trajectory consists of several B$\acute{\text{e}}$zier curves and vehicles will perform multi-stage motion control to track the trajectory. In this paper, the cubic B$\acute{\text{e}}$zier curve with four control points is chosen for single-segment trajectory planning.

\subsection{Multi-stage Motion Control}

As given in Rule \ref{rule1} and Rule \ref{rule2}, vehicles arrive at the $\text{P}_{\text{re}}$s at each $t_i$ in $\mathbb{T}=\{ t_i | t_i=iT, i=0,1,2,3...\}$ in RCS. B$\acute{\text{e}}$zier curves are generated for the vehicles to pass through the corresponding $\text{P}_{\text{ro}}$s in GCS. The key is to calculate longitudinal and lateral control inputs for the vehicles to arrive at the desired position at desired time. The trajectory in GCS passes through a sequence of $\text{P}_{\text{ro}}$s and the inaccuracy during the trajectory following process may accumulate by time. To prevent severe following error which may lead to collision, a multi-stage motion control framework is proposed and vehicles will replan their control inputs after desired time intervals.

\subsubsection{Vehicle Model}

The vehicle model used in this paper is the bicycle model. The center of the rear axle is chosen to represent the position of the vehicle, and the coordinate in GCS is $(x^\mathrm{r},y^\mathrm{r})$. The yaw angle and the steer angle are represented as $\theta$ and $\delta$ respectively. $L$ represents the wheelbase of the vehicle. The state variable $\emph{\textbf{z}}^\mathrm{v}$ of the vehicle contains $x^\mathrm{v}$, $y^\mathrm{v}$, $v$ and $\theta$, where $v$ represents the velocity of the vehicle. The control input for the vehicle model contains the acceleration $a$ and the steer angle $\delta$, and the state is calculated as:
\begin{eqnarray}
&\dot{\emph{\textbf{z}}^\mathrm{v}}=
\begin{bmatrix} v\sin(\theta) \\ v\cos(\theta) \\ a \\ \frac{v}{L}\tan(\delta) \end{bmatrix},
&\emph{\textbf{z}}^\mathrm{v}=
\begin{bmatrix} x^\mathrm{v} \\ y^\mathrm{v} \\ v \\ \theta \end{bmatrix}.
\end{eqnarray}

\subsubsection{Lateral Control}

For the lateral control, a linear feedback preview controller is designed. The vehicle looks ahead to the preview point $P^\mathrm{l}$ and the closest point $P^\mathrm{c}$ on the trajectory. The controller calculates the angle between the vehicle velocity and the desired velocity on $P^\mathrm{l}$, and the distance between the vehicle and $P^\mathrm{c}$ as feedbacks. The detail of the lateral controller design can be found in~\cite{ge2020numerically}.

\subsubsection{Optimal Longitudinal Control With Path Constraints}
The discretized longitudinal state equation is given as:

\begin{eqnarray}
\begin{cases}
&s(k+1)=s(k)+v(k)\Delta t,\\
&v(k+1)=v(k)+a(k)\Delta t,
\end{cases}
\end{eqnarray}
where $s(k)$, $v(k)$ and $a(k)$ represent the longitudinal position, speed and acceleration respectively. The time is discretized by sample interval $\text{d}t$. The distance that the vehicle will travel from the starting point to the final point is denoted as $S_t$. Since the vehicle moving in RCS should pass a sequence of $\text{P}_{\text{re}}$s, $S_t$ consists of series of segments, whose length are denoted as $S_1$, $S_2$, ..., $S_{N^t}$, where $N^t$ is the number of segments. The time interval for the vehicle to cover each segment is $T$. Taking the control energy of the whole control process as the cost, the discretized longitudinal control can be described as:
\begin{alignat}{2}
\min\quad & \sum_{k=0}^{N^tN^k-1} a^2(k), &{}& \label{eqn - lon}\\
\mbox{s.t.}\quad
&s(0)=0,\notag \\
&s(k_i)=\sum_{n=1}^{i}S_n,\ k_i=iN^k,\ i=1,2,3,..., N^t,\notag \\
&v(0)=v(N^tN^k)=v_\mathrm{F},\notag\\
&v_{\text{min}}\leq v\leq v_{\text{max}},\notag\\
&a_{\text{min}}\leq a\leq a_{\text{max}},\notag
\end{alignat}
where $N^k=\frac{T}{\text{d}t}$ is the steps in one segment, $a(k)$ is the longitudinal control input (acceleration) of the vehicle at step~$k$, $v(k)$ is the longitudinal speed, $v_\mathrm{F}$ is the desired longitudinal speed of the formation, and $v_{\text{min}}$, $v_{\text{max}}$, $a_{\text{min}}$ and $a_{\text{max}}$ are the bounds of speed and acceleration.

The optimal control problem in (\ref{eqn - lon}) is solved and the sequence of control input is calculated to guide the vehicle to travel desired distance at desired time. Since the linear feedback preview controller is designed for lateral control, the inaccuracy of lateral motion may cause deviation for longitudinal motion. In order to resolve the accumulated inaccuracy, the optimal problem is reformed and solved every time the vehicle completes the following of one segment at time $t=iT (i=1,2,3,...,N^t)$.

%
\section{Simulation and Results}
\label{simu}
%

In this section, the simulation is conducted and the results are analyzed. The simulation is implemented with MATLAB 2017b and SUMO 0.32.0~\cite{lopez2018microscopic} on a personal computer with CPU Intel CORE i7-8700@3.2GHz. The optimal control problem in (\ref{eqn - lon}) is solved with the help of the multiple-phase optimal control problem solver GPOPS-II~\cite{patterson2014gpops}. The average travel time and fuel consumption of vehicles is analyzed to evaluate the performance. The parameters chosen for this simulation is presented in Table~\ref{para}.

\begin{table}[htbp]
\centering
\caption{Simulation parameters}
\label{para}
\begin{tabular}{lll}
\toprule
Safe one-lane following gap                         &   $d_\mathrm{g}$          & $15\,\mathrm{m}$ \\ 
Formation switching cycle                          &   $T$              & $5\,\mathrm{s}$ \\ 
Desired speed in formation                          &   $v_\mathrm{F}$              & $28.8\,\mathrm{m/s}$\\ 
Minimum speed of vehicle                          &   $v_{\text{min}}$              & $0\,\mathrm{m/s}$ \\ 
Maximum speed of vehicle                          &   $v_{\text{max}}$              & $33.3\,\mathrm{m/s}$\\ 
Minimum acceleration of vehicle              &   $a_{\text{min}}$              & $-10\,\mathrm{m/s^2}$ \\ 
Maximum acceleration of vehicle                 &   $a_{\text{max}}$          & $5\,\mathrm{m/s^2}$ \\ 
Minimum steering angle of vehicle                 &   ${\delta}_{\text{min}}$              & $-40\,\mathrm{^{\circ}}$ \\ 
Maximum steering angle of vehicle                   &   ${\delta}_{\text{max}}$           & $40\,\mathrm{^{\circ}}$ \\ 
\bottomrule  
\end{tabular}
\end{table}

\begin{figure}[tb]
\begin{center}
    \subfigure[Number of lanes changes from three to one]{
        \includegraphics[width=0.95\linewidth]{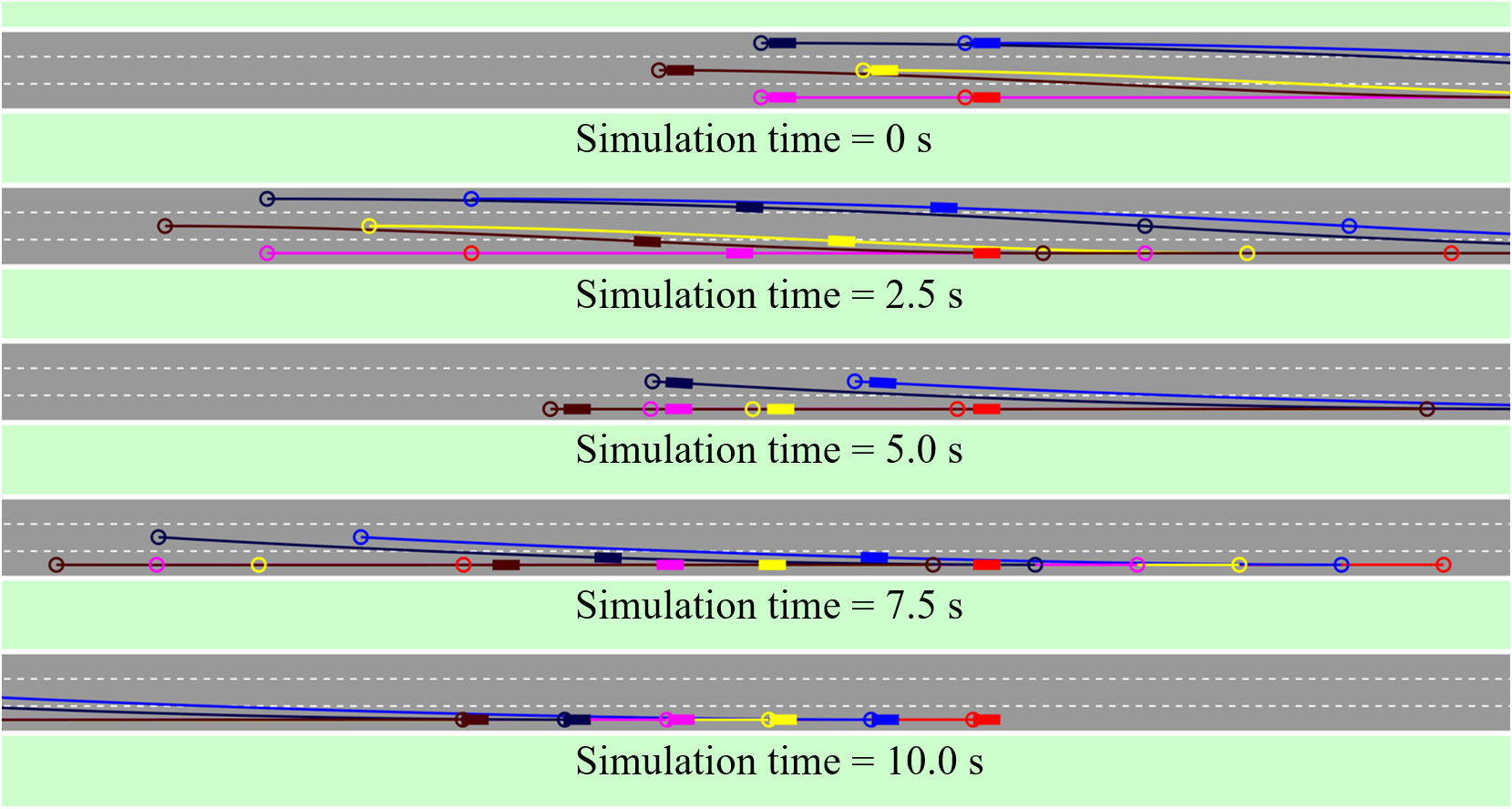}
        \label{3-1}}
    \subfigure[Number of lanes changes from three to two]{
        \includegraphics[width=0.95\linewidth]{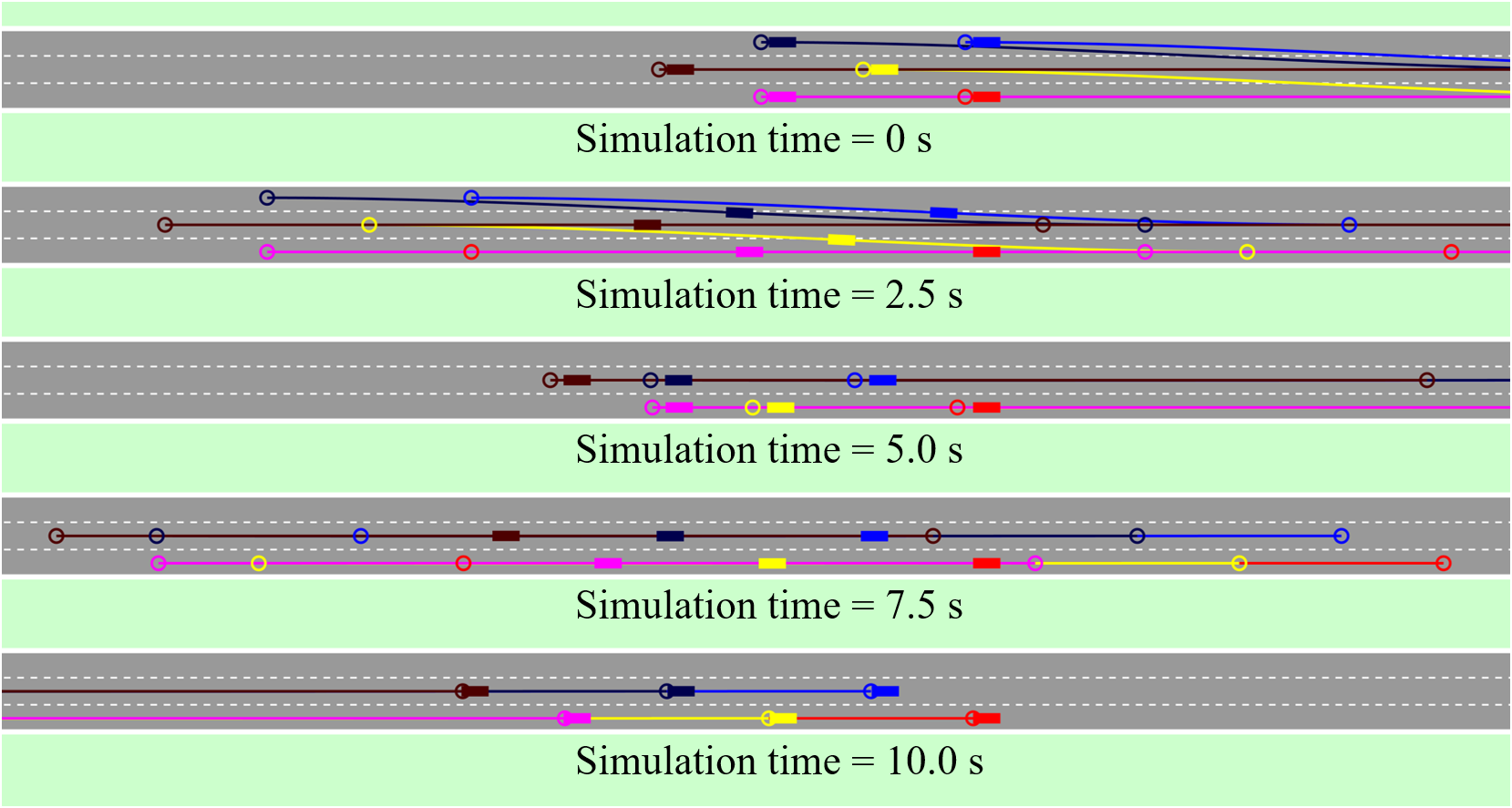}
        \label{3-2}}
     \subfigure[Number of lanes changes from three to four]{
        \includegraphics[width=0.95\linewidth]{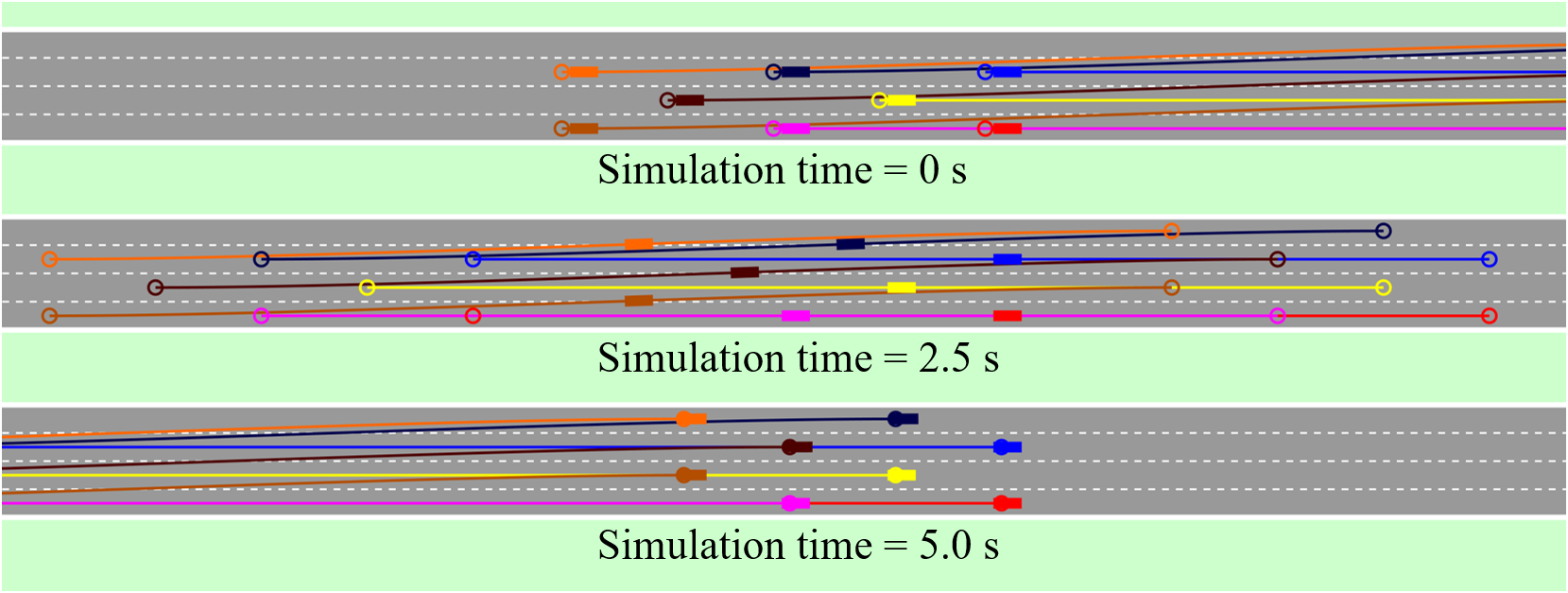}
        \label{3-4}}
    \caption{Formation switching process in different scenarios. Rectangles represent vehicles and the circles and curves with the same color represent the $\text{P}_{\text{ro}}$s and trajectories of the vehicles. The first two cases take two time cycles to complete the process and the last case takes one time cycle. The full-version video of the simulation is available at: \protect\\ {\color{blue}https://github.com/cmc623/Multi-lane-formation-control}}
    \label{process}
\end{center}
\end{figure}

Case study is firstly conducted to verify the function of the proposed method. Three scenarios are chosen for case study: (1) the number of lanes changes from three to one; (2) the number of lanes changes from three to two; (3) the number of lanes changes from three to four. The formation switching process is shown in Fig.~\ref{process}, where the initial interlaced formations switch their structure to adapt to the changing scenarios. It indicates that the proposed method is able to handle different scenarios and can switch the formation smoothly with continuous and smooth trajectories.

\begin{figure}[t]
\begin{center}
    \includegraphics[scale = 0.18]{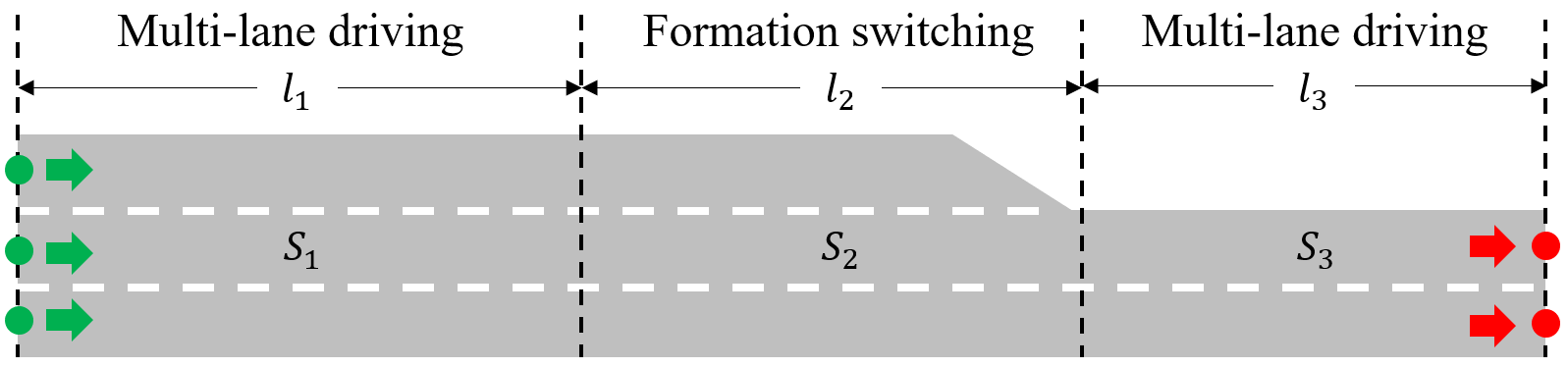}
    \caption{The lane-drop bottleneck scenario. Vehicles travel from left to right and pass $S_1$, $S_2$ and $S_3$ in order.}
    \label{scenariopic}
\vspace{-4mm}
\end{center}
\end{figure}

The lane-drop bottleneck from three lanes to two lanes are chosen as the scenario of the comparative simulation, as shown in Fig.~\ref{scenariopic}. The road segment is divided into three parts: the upstream multi-lane driving part $S_1$, the formation switching part $S_2$, and the downstream multi-lane driving part $S_3$. The length of them are $l_1+l_2=1000\,\mathrm{m}$ and $l_3=200\,\mathrm{m}$. The beginning point of $S_2$ is determined according to the result of motion planning of each single formation. The number of lanes of $S_1$ and $S_2$ are three, and that of $S_3$ is two. Vehicles are generated at the beginning of $S_1$ (the green points) and try to form standard formation when driving on $S_1$. The formation calculates the time cycle it needs to switch to the two-lane structure and thus get the length of $S_2$ ($l_2$). Formation switching is performed on $S_2$ and the third lane (on the top) is cleared before the most forward vehicle of the formation arrives at the beginning of $S_3$. Vehicles leave the road segment at the end of $S_3$ (the red points).

The default driving model of SUMO is adopted as reference. The longitudinal car following model is Intelligent Driving Model (IDM) and the lateral lane change model is the default strategic lane changing~\cite{krauss1998microscopic}. The simulation is conducted under different input traffic volume changing from $250\,\mathrm{veh/(hour\cdot lane)}$ to $2000\,\mathrm{veh/(hour\cdot lane)}$. The simulation time under each traffic volume is $600\,\mathrm{s}$. The average travel time of vehicles on the whole 1200-meter road segment and the converted fuel consumption per $100\,\mathrm{km}$ are chosen to compare the performance of the two methods. The Akcelik’s fuel consumption model is utilized to calculate transient fuel consumption of vehicles \cite{akcelik1989efficiency}.

\begin{figure}[t]
\begin{center}
    \subfigure[$1000\,\mathrm{v/(h\cdot l)}$]{
        \includegraphics[scale = 0.17]{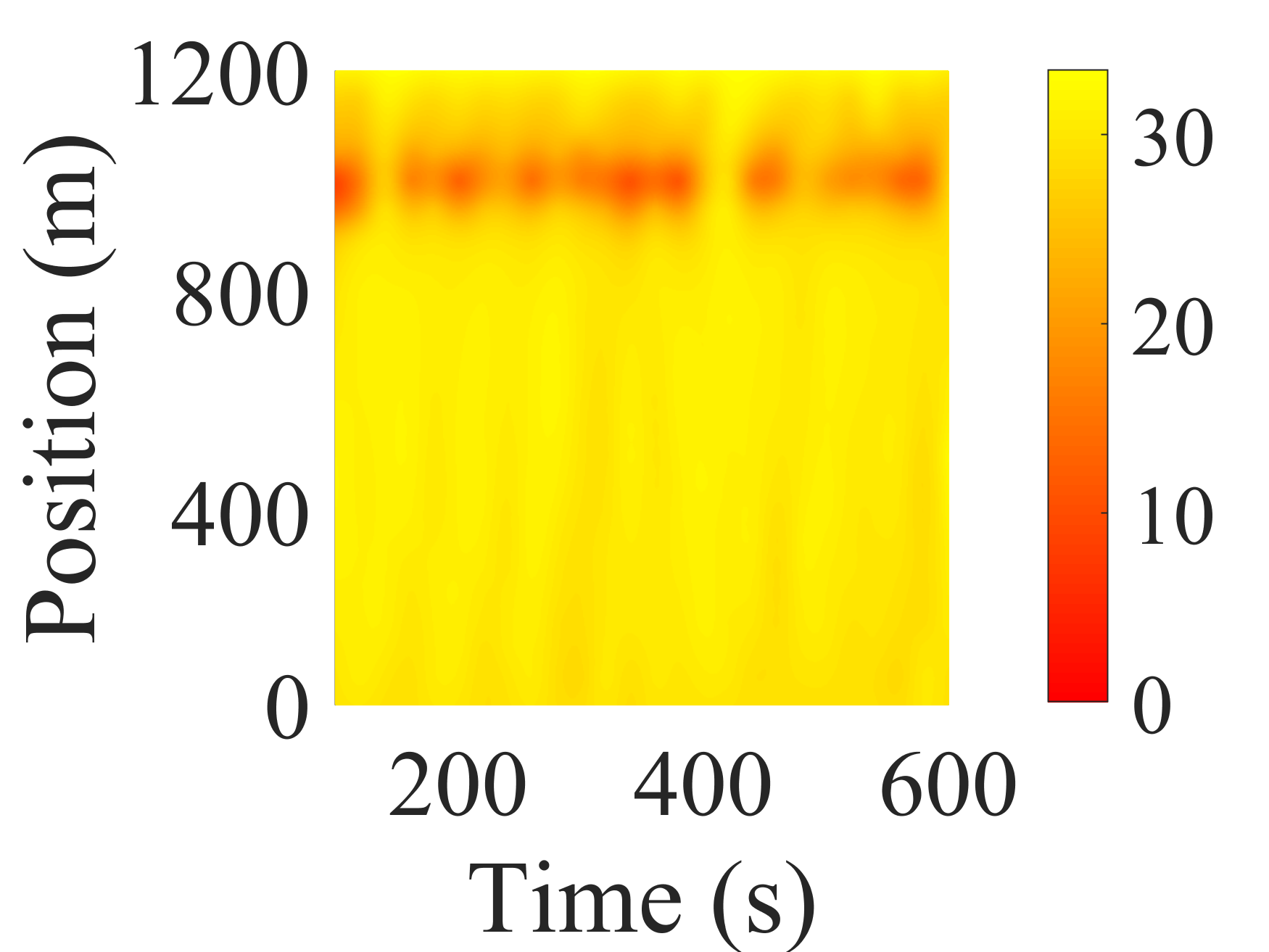}
        \label{ref2}}
\hspace{-3mm}
    \subfigure[$1500\,\mathrm{v/(h\cdot l)}$]{
        \includegraphics[scale = 0.17]{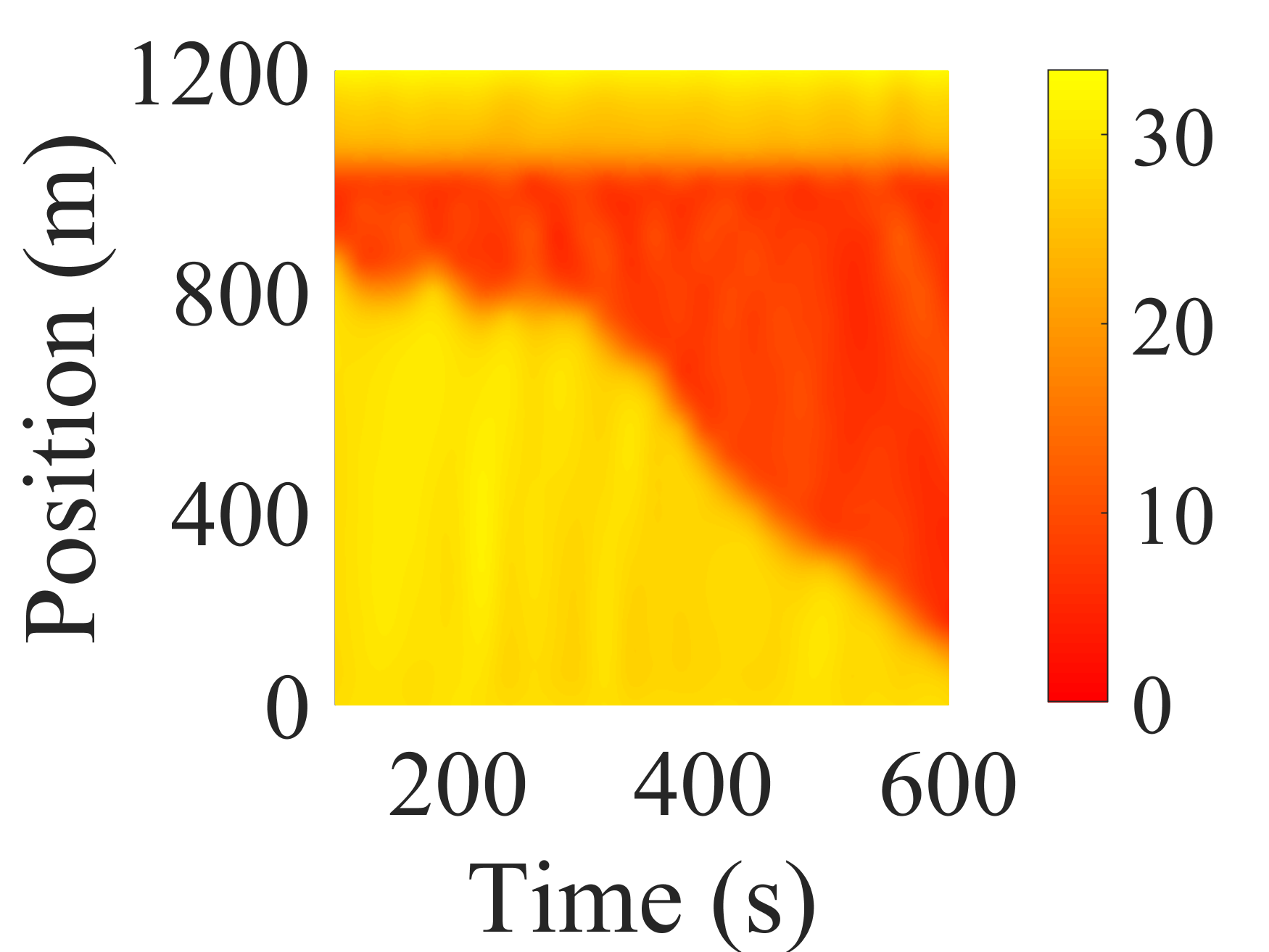}
        \label{ref3}}
\hspace{-3mm}
    \subfigure[$2000\,\mathrm{v/(h\cdot l)}$]{
        \includegraphics[scale = 0.17]{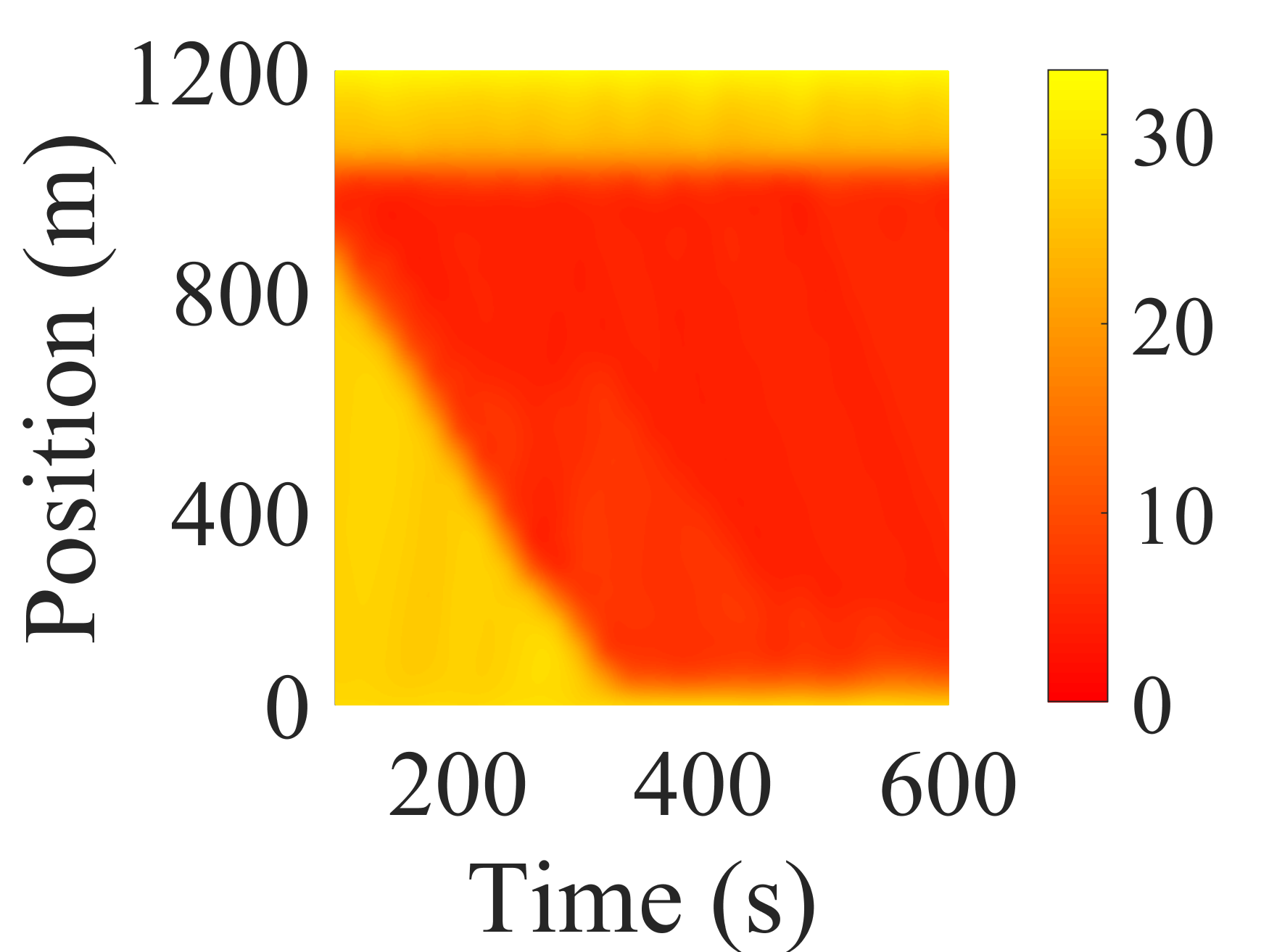}
        \label{ref4}}
    \subfigure[$1000\,\mathrm{v/(h\cdot l)}$]{
        \includegraphics[scale = 0.17]{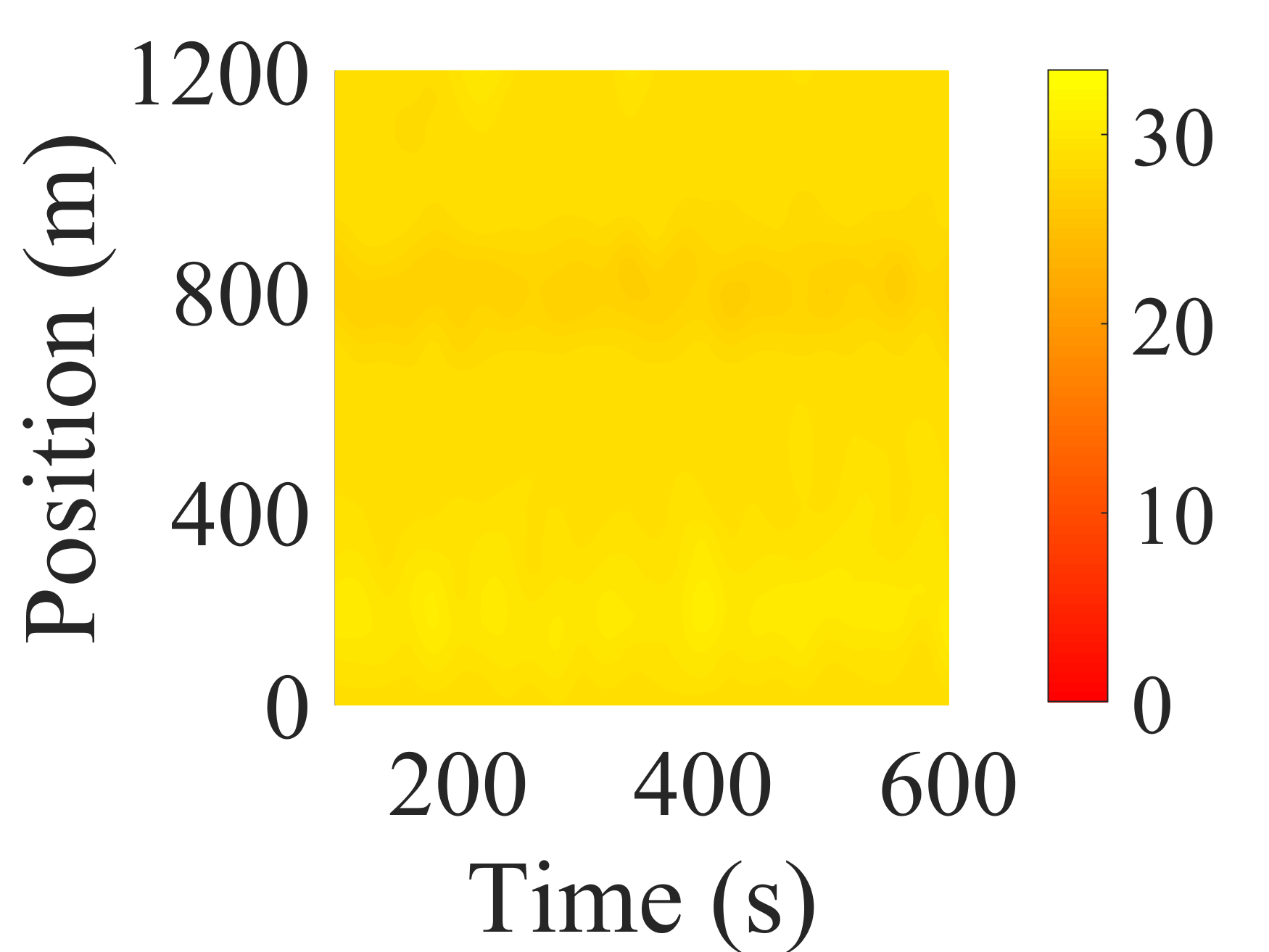}
        \label{fc2}}
\hspace{-3mm}
    \subfigure[$1500\,\mathrm{v/(h\cdot l)}$]{
        \includegraphics[scale = 0.17]{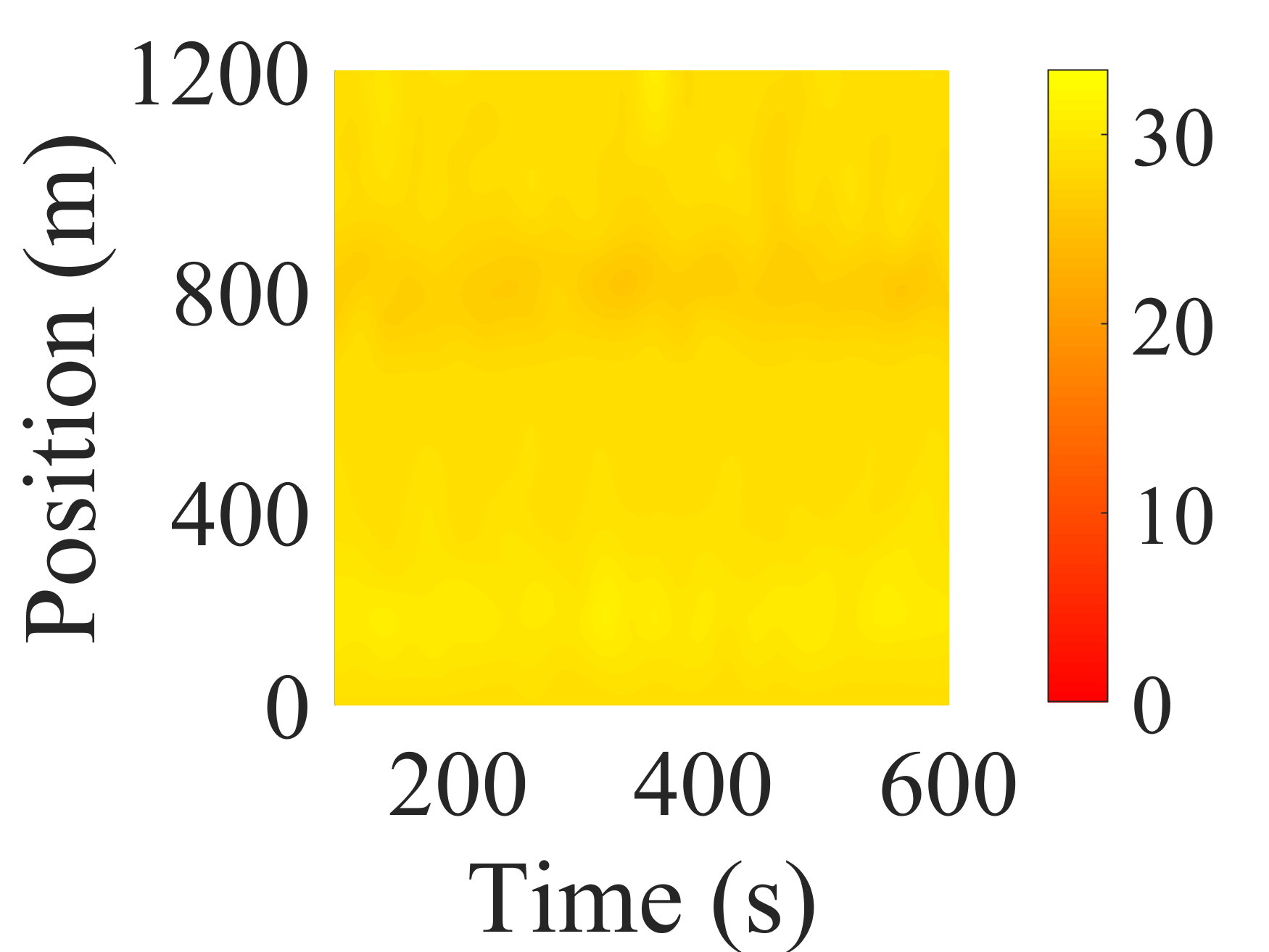}
        \label{fc3}}
\hspace{-3mm}
    \subfigure[$2000\,\mathrm{v/(h\cdot l)}$]{
        \includegraphics[scale = 0.17]{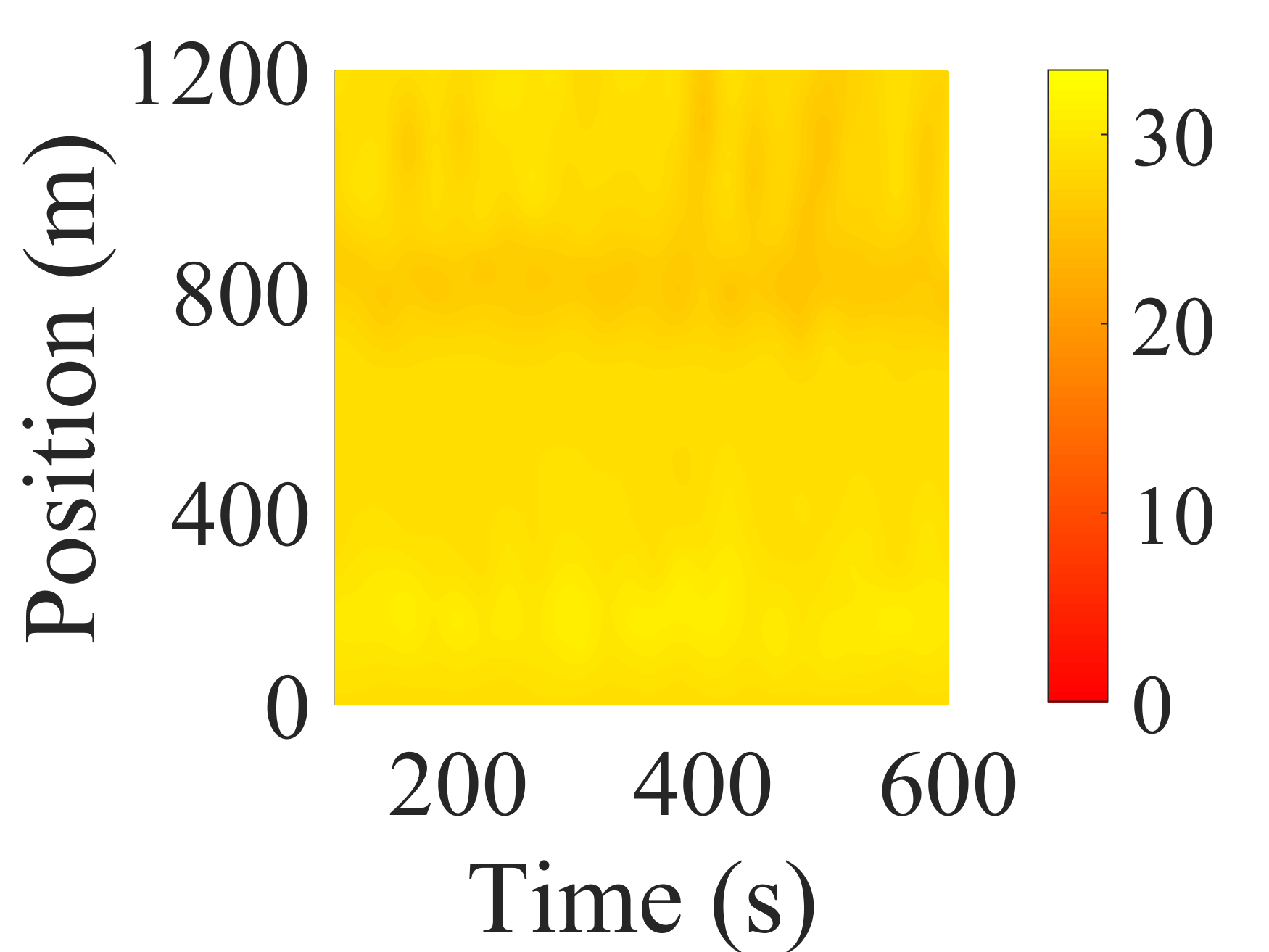}
        \label{fc4}}
    \caption{Heatmaps of the results. (a), (b) and (c) are results without FC. The yellow parts represent that vehicles pass these regions with high speed. It can be seen that congestion is formed under the above three traffic volumes, and the higher the volume is, the severer the congestion is. (d), (e) and (f) are results with FC. It can be seen that under the three traffic volumes, although some slight disturbance of speed happens near the merging point, no severe congestion is formed.}
    \label{heatmap}
\end{center}
\end{figure}

Heatmaps are presented in Fig.~\ref{heatmap} to show the speed distribution among vehicles on time and space dimensions under different traffic input volume. When vehicles driving without FC, congestion may happen and spread upstream from the lane-drop bottleneck point ($1000\,\mathrm{m}$). When FC is conducted, the congestion is significantly reduced and vehicles tend to maintain a constant formation speed ($28.8\,\mathrm{m/s}$).

\begin{figure}[tb]
\begin{center}
    \subfigure[Simulation with FC]{
        \includegraphics[scale = 0.18]{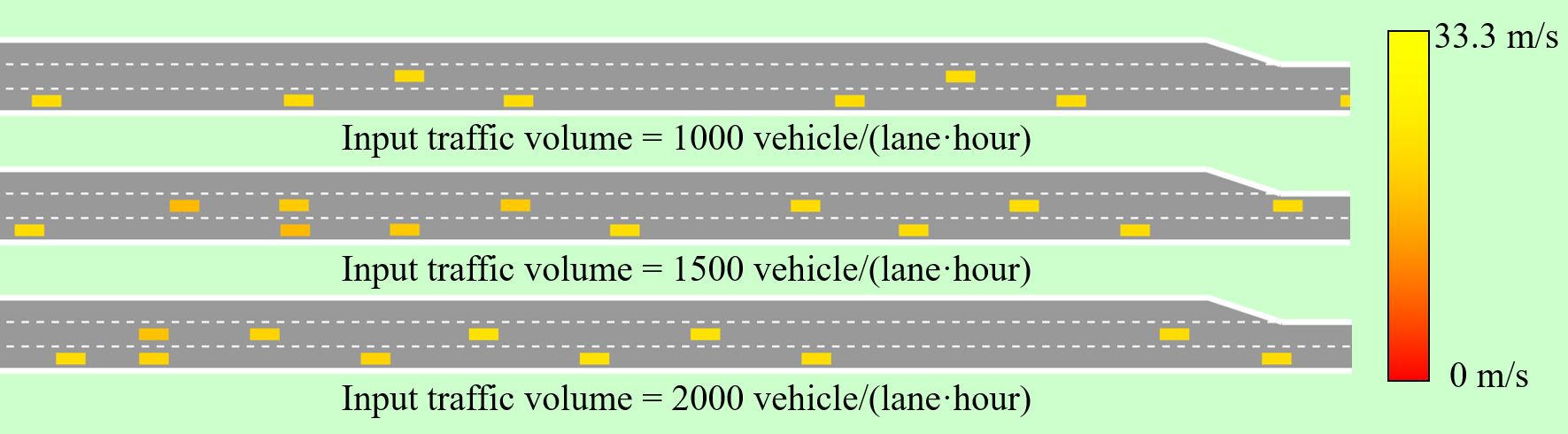}
        \label{reFRC}}
    \subfigure[Simulation without FC]{
        \includegraphics[scale = 0.18]{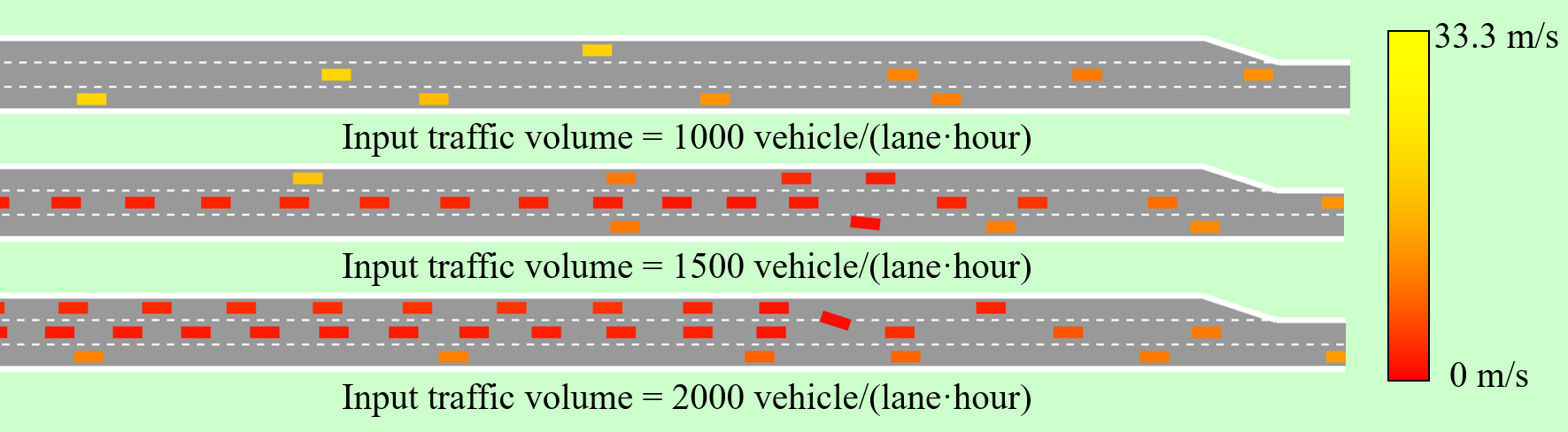}
        \label{reREF}}
    \caption{Snapshots of the simulation process at simulation time $600\,\mathrm{s}$. Different colors represent different speed of vehicles. As shown on the right, the yellower the color is, the higher the speed is, and the redder the color is, the slower the speed is. The full-version video of the simulation is available at: {\color{blue}https://github.com/cmc623/Multi-lane-formation-control}}
    \label{snapshots}
\vspace{-3mm}
\end{center}
\end{figure}

The snapshots of the simulation process are provided in Fig.~\ref{snapshots}. The figures are taken at the fixed simulation time $600\,\mathrm{s}$ under each input traffic volume. The color of the vehicles represents their speed. From the snapshots we can see that the proposed FC method reasonably distributes vehicles on the lanes and guides them to drive with the expected speed ($28.8\,\mathrm{m/s}$) and avoids congestion at the lane-drop bottleneck. Although some vehicles can achieve higher speed when not using FC, the performance at the bottleneck is poor and severe congestion is formed under high traffic volume.

\begin{figure}[tb]
\begin{center}
    \subfigure[Average travel time of vehicles]{
        \includegraphics[scale = 0.55]{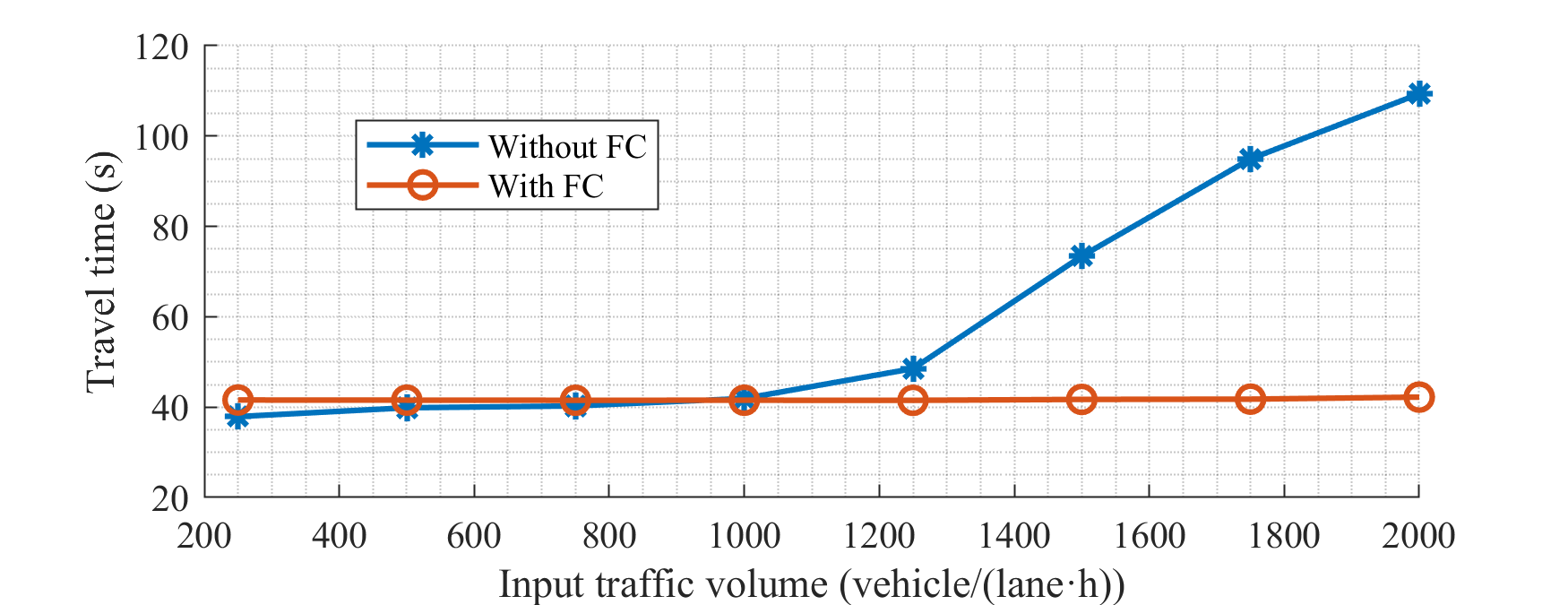}
        \label{att}}
    \subfigure[Average fuel consumption of vehicles]{
        \includegraphics[scale = 0.55]{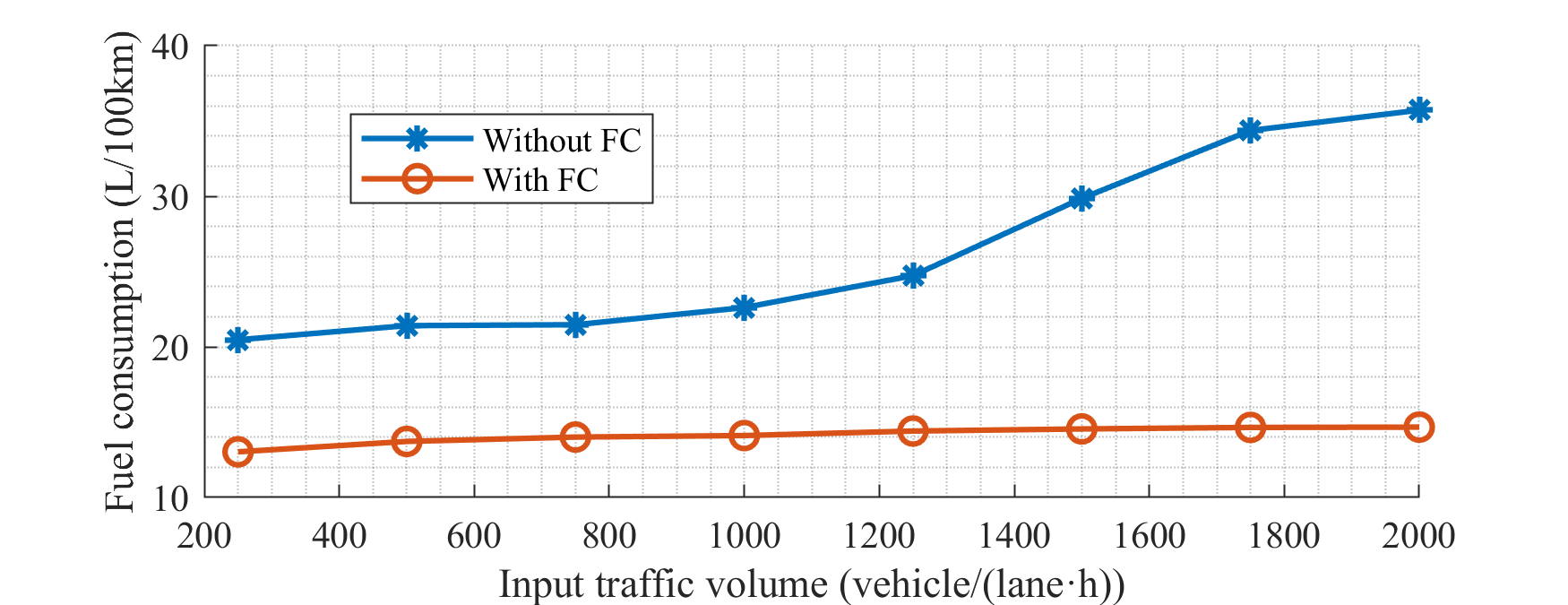}
        \label{afc}}
    \caption{Numerical results of the simulation in the lane-drop bottleneck scenario. The orange lines show the result with FC and the blue lines show the result without FC.}
    \label{resultpic}
    \vspace{-3mm}
\end{center}
\end{figure}

Numerical results are shown in Fig.~\ref{resultpic}. It indicates that vehicles pass the whole road segment faster when not using FC with the input traffic volume lower than $1000\,\mathrm{veh/(hour\cdot lane)}$. It's because there is no or slight congestion near the bottleneck and the expected speed of vehicles is higher when not using FC. When the input traffic volume gets higher, the congestion becomes severer, the average speed of vehicles becomes lower, and the average travel time gets higher. The average fuel consumption without FC also goes higher when the traffic volume gets higher, and is always greater than the fuel consumption of FC method. It is noticeable that both the travel time and the fuel consumption of the FC method keep almost unchanged under different traffic volume, because the input traffic volume hasn't reach the theoretical maximum congestion-free volume, which is around $2500\,\mathrm{veh/(hour\cdot lane)}$ and can be calculated by the following gap and desired formation speed.

%
\section{Conclusions}
\label{conc}
%

This paper proposes a formation control method for multiple vehicles driving on multi-lane roads. The bi-level planning framework is proposed to smoothly and effectively switch the structure of formation in different scenarios. Relative coordinate system is built for relative motion planning and conflict resolution in the upper level. Multi-stage trajectory planning and tracking are designed to follow the conflict-free paths in the lower level. Case study and numerical results of the simulation indicate that:
\begin{enumerate}
\item the proposed FC method is able to smoothly switch the structure of formation to adapt to different traffic scenarios.
\item the proposed FC method can improve traffic efficiency (travel time) under high traffic volume because no severe congestion is formed and vehicles tend to travel with a stable speed.
\item the proposed FC method can improve fuel economy under all the tested traffic volume because vehicles don't accelerate and decelerate frequently.
\end{enumerate}

The future directions of this research include extending the proposed FC framework to ramp merging and multi-lane intersection scenarios, and the stability analysis of the stable-driving formation with information or control disturbance.

\bibliographystyle{IEEEtran}
\bibliography{thesis}

\end{document}